\documentclass{bull-l}

\usepackage{amssymb,upref,verbatim}
\newcommand{\bbN}{{\mathbb{N}}}
\newcommand{\bbR}{{\mathbb{R}}}

\newcommand{\bbZ}{{\mathbb{Z}}}
\newcommand{\bbC}{{\mathbb{C}}}

\newcommand{\calK}{{\mathcal K}}

\newcommand{\calS}{{\mathcal S}}
\newcommand{\calT}{{\mathcal T}}

\newcommand{\calW}{{\mathcal W}}
\newcommand{\calY}{{\mathcal Y}}

\newcommand{\no}{\nonumber}
\newcommand{\lb}{\label}
\newcommand{\f}{\frac}
\newcommand{\ul}{\underline}

\newcommand{\ti}{\widetilde}

\newcommand{\s}{{n+1}}
\newcommand{\R}{R_{2n+2}}
\newcommand{\F}{F_n}
\newcommand{\tF}{\tilde{F}_{n+1}}
\newcommand{\hF}{\hat{F}_{\hat{n}}}
\newcommand{\hR}{\hat{R}_{2n-2s+4}}

\newcommand{\bi}{\bibitem}

\renewcommand{\Re}{\operatorname{Re}}
\renewcommand{\Im}{\operatorname{Im}}
\newcommand{\grad}{\operatorname{grad}}
\newcommand{\df}{\operatorname{def}}
\newcommand{\ord}{\operatorname{ord}}
\renewcommand{\atop}[2]{\genfrac{}{}{0pt}{}{#1}{#2}}
\newcommand{\e}{\hbox{\rm e}}

\DeclareMathOperator{\tr}{tr}

\allowdisplaybreaks

\numberwithin{equation}{subsection}

\newtheorem{theorem}{Theorem}[section]

\newtheorem{corollary}[theorem]{Corollary}

\theoremstyle{definition}
\newtheorem{definition}[theorem]{Definition}

\theoremstyle{remark}
\newtheorem{remark}[theorem]{Remark}

\begin{document}
\title[Elliptic Solutions of Soliton Hierarchies] {Elliptic
Algebro-Geometric Solutions of the KdV and AKNS
Hierarchies -- An
Analytic Approach}
\author{Fritz Gesztesy and Rudi Weikard}
\address{Department of Mathematics,
University of Missouri,
Columbia, MO 65211, USA}
\email{fritz@math.missouri.edu}
\urladdr{http://www.math.missouri.edu/people/faculty/fgesztesypt.html}
\address{Department of Mathematics,
University of Alabama at Birmingham, Birmingham, AL 35294-1170, USA}
\email{rudi@math.uab.edu}
\urladdr{http://www.math.uab.edu/rudi}

\thanks{Research supported in part by the
US National Science Foundation under Grant Nos. DMS-9401816 and
DMS-9623121}
\date{May 7, 1998}

\begin{abstract}
We provide an overview of elliptic algebro-geometric
solutions of the
KdV and AKNS hierarchies, with special emphasis on
Floquet theoretic and
spectral theoretic methods. Our treatment includes an
effective
characterization of all stationary elliptic KdV and AKNS
solutions based
on a theory developed by Hermite and Picard.
\end{abstract}
\maketitle
\section{Introduction.} \label{introduction}

The story of J. Scott Russell chasing a soliton for a
mile or two along
the Edinburgh-Glasgow channel in 1834 has been told many
times. It is
the starting point of more than 160 years of an exciting
history
embracing a variety of deep mathematical ideas ranging
from applied
mathematics to algebraic geometry, Lie groups, and
differential
geometry. In this article we want to tell a certain
aspect of this
story, predominantly from an analytic point of view
with special
emphasis on Floquet theoretic and spectral theoretic
methods. Other
aspects of this story have recently been told by
Lax \cite{La96}, Palais
\cite{Pa97}, and Terng and Uhlenbeck \cite{TU97}.

We are interested in the class of evolution equations
which permit being
cast into the form $L_t=[P,L]$, where $(P,L)$ is a pair of
operators, a
so called Lax pair, and $[P,L]$ denotes their commutator.
Typically, for
a fixed operator $L$, there is a sequence of operators
$P$ such that
$L_t=[P,L]$ defines an evolution equation. Hence we
are actually
concerned with hierarchies of evolution equations.
Stationary solutions
of such equations, corresponding to commuting operators
$P$ and $L$, are
related to algebraic curves and are therefore called
algebro-geometric
solutions. The stationary solutions of higher-order
equations in the
hierarchy play a decisive role in the study of the Cauchy
problem for
the lower-order equations in the hierarchy. Therefore,
and because of
the connection to algebraic geometry, the stationary
problem has drawn
considerable attention.

The central object of this article is the class of
elliptic
algebro-geometric solutions of the KdV and AKNS
hierarchies
(cf.~Section~\ref{laxp}). In a sense, this class
represents a natural
generalization of the class of soliton solutions and
enjoys a much
richer structure due to its connections to algebraic
geometry inherent
in its construction. However, while frequently the
algebraic aspects of
this construction dominate the stage, we will purposely
portray a
different analytical view often neglected (and to
some extent forgotten)
in this context. Consequently, we will focus in the
following on the
interplay between spectral (and Floquet) theoretic
properties of the Lax
pairs (cf.~Section~\ref{laxp}) defining the integrable
evolution
equations in question on the one hand, and the
construction and
properties of the underlying compact Riemann surface a
ssociated with
algebro-geometric solutions on the other.

In our quest to characterize the class of elliptic
algebro-geometric
solutions of soliton equations in an effective manner,
we rely heavily
on a marvelous theory developed by Fuchs, Halphen,
Hermite,
Mittag-Leffler, and especially, Picard. We have chosen
to provide a
rather extensive bibliography concerning this classical
work since some
of the cornerstones of this theory seem to have been
forgotten, and at
times appear to be independently rediscovered
(cf.~Subsection~\ref{lde}
and the bibliographical remarks at the end of
Remark~\ref{R5.2}). In
particular, Picard's theorem (Theorem~\ref{picard}), a
key in the
aforementioned characterization problem, apparently had
not been used
in the extensive body of literature surrounding elliptic
algebro-geometric solutions of various hierarchies
of soliton equations.
Picard's theorem suggests one consider the independent
variable of the
differential equation in question as a complex variable,
and study
the consequences of assuming the existence of a
meromorphic fundamental
system of solutions. This led to the discovery of a
connection between
the existence of such a meromorphic fundamental system
of solutions of a
linear differential equation $Ly=zy$ for all values of
$z$ and
algebro-geometric properties of $L.$ In particular,
it connects the
existence of a meromorphic fundamental system of
solutions of $Ly=zy$ to
the integrability of the nonlinear equations associated
with $L$ via the
Lax pair formalism, as we demonstrated in the KdV and
AKNS cases
(cf.~Theorems \ref{t3.12} and \ref{t5.4}). As it turns
out, Picard was
describing solutions of differential equations with
elliptic
coefficients, which represent simultaneous Floquet
solutions with
respect to all fundamental periods of the underlying
period lattice of
the torus in question. Hence Floquet theory, and
consequently, spectral
theory, naturally enters when analyzing Lax pairs for
completely
integrable evolution equations and their elliptic
algebro-geometric
solutions.

Section~\ref{history} provides an introduction into
the KdV hierarchy
and its elliptic algebro-geometric solutions, reviews
the necessary
Floquet and spectral theoretic background for Hill
(Lax) operators,
presents Picard's theorem on linear differential equations
with elliptic
coefficients, and especially, provides a historical
perspective of the
subject. Section~\ref{agkdvp} is devoted to the KdV
hierarchy and its
stationary solutions. The case of rational, periodic,
and elliptic
stationary KdV solutions is described in detail and
the role of
meromorphic fundamental systems of solutions for the
corresponding Lax
operator associated with stationary KdV solutions is
underscored. This
section culminates in an explicit characterization of
all elliptic,
simply periodic, and rational stationary KdV solutions
(see
\cite{GW95d}, \cite{GW96}, \cite{We98a} for the original
results). Our
final Section~\ref{agaknsp} describes analogous results
for the AKNS
hierarchy, in particular, it contains an effective
characterization of
all elliptic algebro-geometric AKNS solutions
(cf.~\cite{GW98} for the
original proof). We hope the enormous bibliography
at the end (still
necessarily incomplete), will lose some of its
intimidating character
once the reader begins to appreciate the large body of
knowledge amassed
by some of the giants in the field, starting with Hermite.

Finally we remark that connections between completely
integrable
sytems and Seiberg-Witten theory via elliptic
Calogero-Moser-type
models, have recently led to a strong resurgence of the
field of elliptic
solutions of soliton equations. It seems difficult to
keep up with
all the current activities in the corresponding preprint
archives.
Hence we refer, for instance, to Babelon and Talon
\cite{BT97},
Donagi and Markman \cite{DM96}, Donagi and Witten
\cite{DW96},
Itoyama and Morozov \cite{IM96}, Krichever \cite{Kr98},
Krichever, Babelon, Billey, and Talon \cite{KBBT95},
Krichever and
Phong \cite{KP97}, Krichever, Wiegmann, and Zabrodin
\cite{KWZ98}, Krichever and Zabrodin \cite{KZ95},
Kuznetsov, Nijhoff,
and Sklyanin \cite{KNS97}, Levin and Olshanetsky
\cite{LO97},
Marshakov \cite{Ma97}, and Vaninsky \cite{Va97}),
from which the interested reader can easily find further
sources.

\section{Background and some History.} \label{history}
\subsection{The Korteweg-de Vries Equation.} \label{kdv}

\setcounter{equation}{0}

In the first few decades after their discovery, solitary
waves were
considered, for instance, by Stokes, Boussinesq, and Lord
Rayleigh. But
the most far reaching, and in a sense, lasting contribution
was made in 1895 by Korteweg and
de Vries \cite{KV95}, who deduced their
celebrated equation, which may be cast in the form
\begin{equation}
q_t=\frac14 q_{xxx}+\frac32 qq_x. \lb{kdveq}
\end{equation}
(However, as recently alluded to by Pego \cite{Pe98},
Boussinesq
originally noted a system of equations equivalent
to \eqref{kdveq} and
used it to study solitary waves in the 1870's.) In
1960 Gardner and
Morikawa \cite{GM60} used the KdV equation to describe
collisionless-plasma magnetohydromagnetic waves. Since
then the KdV
equation has been rederived again and again in
different contexts as a
model equation describing a considerable variety of
physical phenomena,
and can now be considered one of the basic equations
in mathematical
physics. Even more important is the fact that this
development triggered
the examination of a whole class of other nonlinear
evolution equations,
some of which are of great physical relevance, such as
the nonlinear
Schr\"odinger equation, the sine-Gordon equation, the
Toda lattice, the
Boussinesq equation, the Kadomtsev-Petviashvili
equation, etc.

But the description of wave phenomena alone would not
have turned the
subject into the kind of industry it is today. In
1955 Fermi, Pasta, and
Ulam \cite{FPU65} studied a system of nonlinear
oscillators which may be
viewed as a discretized version of the KdV equation
on a finite interval
with periodic boundary conditions. To their surprise
they found that
energy shows little tendency toward equipartition
among the degrees of
freedom. Later Kruskal and Zabusky \cite{ZK65} examined
the KdV equation
numerically and observed the formation of solitary
waves which, using
their words, ``pass through one another without losing
their identity''.
In order to emphasize this particle-like behavior
Kruskal and Zabusky
coined the term soliton.

In 1968 Miura \cite{Mi68} introduced the transformation
\begin{equation} \label{miura}
q(x,t)=z-v(z,x,t)^2-v_x(z,x,t)
\end{equation}
in which $z$ is a (generally complex) spectral parameter.
This
transformation, now known as
Miura's transformation, relates the KdV equation to
a variant of the so
called modified KdV (mKdV) equation,
\begin{equation}
v_t=v_{xxx}-6v^2v_x+6zv_x.
\end{equation}
Since
\eqref{miura} is a Riccati equation for $v,$ it may
be transformed into
the linear equation
$$(L(t)y)(z,x,t)=y''(z,x,t)+q(x,t)y(z,x,t)=zy(z,x,t)$$
by introducing $v(z,x,t)=y'(z,x,t)/y(z,x,t)$ (primes
denote derivatives
with respect to $x$). Gardner, Kruskal, and Miura
\cite{MGK68} showed
that the eigenvalues of the $L^2(\bbR)$-operator
associated with $L(t)$
do not depend on $t,$ that is, they are constants of
the motion under
the KdV flow. This was the starting point for another
seminal work, this
time by Gardner, Greene, Kruskal, and Miura \cite{GGKM67},
in which they
used the inverse scattering method to solve the Cauchy
problem for the
KdV equation with rapidly decaying initial data. This
method, which
represents a nonlinear analog of the Fourier transform
to solve linear
partial differential equations, initially consists of
computing the
scattering data of $L(0),$ then propagating them in
time (which is
simple), and finally reconstructing the potential
$q(x,t)$ in $L(t)$ via
the Gelfand-Levitan, or rather, the Marchenko equation.
Miraculously,
this function $q(x,t)$ is the desired solution of the
Cauchy problem.
See, for instance, Ablowitz and Clarkson \cite{AC91}, Ch.~2,
Asano and
Kato \cite{AS90}, Chs.~5,6, Dodd, Eilbeck, Gibbon, and
Morris
\cite{DEGM88}, Ch.~4, Drazin and Johnson \cite{DJ89},
Ch.~4, Gardner,
Greene, Kruskal, and Miura \cite{GGKM74}, Iliev, Khristov,
and Kirchev
\cite{IKK94}, Ch.~3, Lax \cite{La96}, Marchenko \cite{Ma86},
Ch.~4, and
Palais \cite{Pa97} for more details.

\subsection{Hamiltonian Systems.} \label{hs}

The dynamics of a classical mechanical particle system is
described as a
flow on a symplectic space, called the phase space. The
phase space is
the cotangent bundle of a Riemannian manifold, called the
configuration
space, describing the positions of the particles. The flow
is given by
Hamilton's equation $q_t=\grad_S (H),$ where $\grad_S$ is
a symplectic
gradient and $H$ is the so called Hamiltonian of the system.
If there
are only finitely many, say $n,$ degrees of freedom (as in
a system of
finitely many interacting particles) then $q=(\xi,\eta),$
where $\xi$ is
a vector of local coordinates on the configuration space
(called
generalized coordinates) and $\eta$ is the vector of the
associated
momenta (called generalized momenta). $H$ is then a
function of $\xi,$
$\eta,$ and $t,$ and the symplectic gradient is given by
$$\grad_S=\begin{pmatrix}0& I_n\\-I_n & 0\end{pmatrix}
 \grad_{(\xi,\eta)},$$
where $I_n$ is the $n\times n$ identity matrix. A canonical
transformation is a change of coordinates in phase space
which leaves
the form of Hamilton's equation invariant. Sometimes
there exists a
canonical transformation such that the transformed
Hamiltonian is a
function of the transformed generalized momenta alone
which are then
called action variables while the transformed generalized
coordinates
are called angle variables. The action variables are then
constants of
the motion while the angle variables change linearly
with time. If this
happens the system is called completely integrable.

In 1968 Gardner, Kruskal, and Miura \cite{MGK68} explicitly
constructed
an infinite sequence of constants of the motion providing
the first hint
toward a Hamiltonian structure of the KdV equation. Next,
in 1971,
Faddeev and Zakharov \cite{ZF71}, in an attempt to explain
the unusual
behavior of the KdV equation, showed that it can be viewed
as an
infinite-dimensional completely integrable Hamiltonian
system. Take, for
instance, the Schwartz space $\calS(\bbR)$ as phase space
which can be
viewed as a symplectic manifold. A symplectic gradient on
$\calS(\bbR)$
is given by
$$\grad_S=\frac{\partial}{\partial x} \grad \, ,$$
where $\grad (F)$ denotes the Lagrangian (or variational)
derivative of
$F:\calS(\bbR)\to \bbR.$ More precisely, if $F$ is of the
form
$$
F(q)=\int_{-\infty}^\infty dx \, \ti F(q,q',\dots,q^{(n)}),
$$
where $\ti F:\bbR^{n+1}\to\bbR$ is a polynomial function
without
a constant term, then
$$\grad (F)=\frac{\partial\ti F}{\partial q}
 -\bigg(\frac{\partial\ti F}{\partial q'}\bigg)'+\cdots
 +(-1)^n \bigg(\frac{\partial\ti F}{\partial q^{(n)}}
\bigg)^{(n)}$$
and hence
$$\grad_S (F)=\bigg(\frac{\partial\ti F}{\partial q}\bigg)'-
 \bigg(\frac{\partial\ti F}{\partial q'}\bigg)''+\cdots
 +(-1)^n \bigg(\frac{\partial\ti F}
 {\partial q^{(n)}}\bigg)^{(n+1)}$$
using primes to denote derivatives with respect to $x$.
The Hamiltonian flow
$$\dot q =\grad_S (F)$$
then becomes the nonlinear evolution equation
$$q_t=\bigg(\frac{\partial\ti F}{\partial q}\bigg)'-
 \bigg(\frac{\partial\ti F}{\partial q'}\bigg)''+\cdots
 +(-1)^n \bigg(\frac{\partial\ti F}
 {\partial q^{(n)}}\bigg)^{(n+1)}.$$
In the special case of the KdV flow, the Hamiltonian $H$
is given by
$$
H:\calS(\bbR)\to \bbR, \quad q\mapsto
\frac14\int_{-\infty}^\infty
 dx\,(q(x)^3-\frac12 q'(x)^2),
$$
and Hamilton's equation $q_t=\grad_S (H)$ results in
the KdV equation
\eqref{kdveq}. Analogous considerations apply to the
periodic case
replacing $\int_{-\infty}^\infty dx$ by $\int_0^\Omega
dx,$ with
$\Omega>0$ the fundamental period of $q.$ We refer, for
instance,
to Palais \cite{Pa97} for more information.

\subsection{Lax Pairs and the KdV Hierarchy.} \label{laxp}

Gardner, Greene, Kruskal, and Miura \cite{GGKM67} had
shown that the
$L^2$-spectrum of $L(t)=\partial^2/\partial x^2+q(x,t)$ is
independent of $t$
whenever $q$ is a solution of the KdV equation. This
discovery
inspired Lax
to conjecture that the operators $L(t)$ are all unitarily
equivalent to
one another, that is, $L(0)=U(t)^{-1} L(t) U(t)$ for
some family of
unitary operators $U(t).$ This then led to his
celebrated commutator representation \cite{La68},
$$q_t=L_t=[P_3,L],$$
where $P_3$ denotes the third-order ordinary differential
expression
defined by
$$P_3=\frac14 \frac{\partial^3}{\partial x^3} +\frac32 q
 \frac{\partial}{\partial x}+\frac34 q_x,$$
which governs the time evolution of $U(t)$ according to
$U_t(t)=P_3(t)U(t),$ $U(0)=I.$ Indeed, the latter equation
implies
$(U(t)^{-1})_t=-U(t)^{-1}P_3(t)$ and hence formally,
\begin{equation}
\frac{d}{dt}(U(t)^{-1}L(t)U(t))
=U(t)^{-1}(L_t(t)-[P_3(t),L(t)])U(t)=0 \no
\end{equation}
yields
\begin{equation}
U(t)^{-1}L(t)U(t)=L(0), \quad t\in\bbR. \no
\end{equation}
(Functional analytic arguments guaranteeing self-adjointness
of $L(t)$
and unitarity of $U(t)$ in the Hilbert space $L^2(\bbR;dx)$
can easily
be supplied.) The letters $P$ and $L$ were first used
by Gelfand and
Dickey \cite{GD76} in honor of Peter Lax. Accordingly,
$(P_3,L)$ is
called a Lax pair.

In fact, Lax designed his procedure for a far more general
setting: it
applies to evolution equations for $q$ whenever there is
a one-to-one
correspondence between $q(\cdot,t)$ and a (formally)
symmetric operator
$L(t).$ The goal is then to find a (formally) skew-symmetric
operator
$P(t)$ satisfying $U_t(t)=P(t)U(t),$ where
$L(0)=U(t)^{-1} L(t) U(t).$
In particular, returning to the case where
$L(t)=\partial^2/\partial
x^2+q(x,t),$ Lax showed that one can always find an
odd-order
differential expression $P_{2n+1}(t)$ such that
$[P_{2n+1}(t),L(t)]$ is
an operator of multiplication. The coefficients of
$P_{2n+1}(t)$ are
then differential polynomials in $q(x,t),$ that is,
polynomials of
$q(x,t)$ and its $x$-derivatives. The commutator
$[P_{2n+1},L]$ is then
also a differential polynomial in $q$ and the evolution
equation
\begin{equation}
q_t=[P_{2n+1},L] \lb{lax}
\end{equation}
defines the $n$-th equation of the KdV hierarchy.

More precisely, if $L=\partial^2/\partial x^2+q,$ then
any ordinary
differential expression $P,$ for which $[P,L]$ equals an
operator of
multiplication, is the sum of a polynomial $K\in\bbC[L]$
and an
odd-order differential expression $P_{2n+1}$, that is,
$$P=K(L)+P_{2n+1}.$$
Upon rescaling the time variable, $P_{2n+1}$ can be
assumed to be monic.
Moreover, choosing the polynomial $K$ appropriately, it
may be shown
that
\begin{equation} \label{09122}
P_{2n+1}=\sum_{j=0}^n \left[-\frac12 f_j'
 + f_j \frac{d}{dx} \right] L^{n-j}
\end{equation}

for some integer $n\in\bbN_0(=\bbN\cup\{0\}),$ where the
functions
$f_j$ satisfy the recursion relation
\begin{align}
 f_0 &=1, \no \\
 {f}'_{j+1} &=\frac{1}{4} {f}^{'''}_j+q{f}'_j
 +\frac{1}{2}q'{f}_j, \quad j=0,\dots,n. \label{recrel}
\end{align}
Then
$$[P,L]=[P_{2n+1},L]=\frac{1}{4}{f}^{'''}_n+q{f}'_n
 +\frac{1}{2}q'{f}_n$$
and introducing
$${F_n}(z,x,t)=\sum^{n}_{j=0}{f}_{n-j}(x,t) z^j,$$
this recursion relation becomes
\begin{equation} \label{aplax}
[P_{2n+1},L]=\frac{1}{2}{F_n}^{'''}+2(q-z){F_n}'+q'{F_n}.
\end{equation}
Conversely, if $F_n(z,x)$ is a polynomial in $z$ such that
${F_n}'''+4(q-z) F'_n+2q' F_n$ does not depend on $z,$
then its
coefficients define a differential expression $P_{2n+1}$
such that
equation \eqref{aplax} is satisfied. In particular, if
${F_n}'''+4(q-z)
F'_n+2q'F_n=0,$ then $q$ is a stationary (i.e.,
$t$-independent)
solution of one of the $n$-th equations in the KdV hierarchy.

The first few equations of the KdV hierarchy explicitly read,
\begin{align*}
 q_t &= q_x,  \\
 q_t &= \frac14 q_{xxx}+\frac32 qq_x + c_1 q_x, \no \\
 q_t &= \frac1{16} q_{xxxxx} +\frac58qq_{xxx}+\frac54q_xq_{xx}
 +\frac{15}8q^2q_x +c_1(\frac14 q_{xxx}+\frac32 qq_x)
+ c_2 q_x,
\no \\
 & \text{etc.} \no,
\end{align*}
where the $c_\ell$ denote integration constants encountered in
solving
the recursion \eqref{recrel} for $f_{j+1}.$ Here we ought
to mention
that each of these equations is a completely integrable
Hamiltonian
system and that the sequence of these equations is intimately
related to
the sequence of conservation laws discovered by Gardner,
Kruskal, and
Miura \cite{MGK68}.

However, the Lax method can be further generalized in a
variety of
ways: If one considers
instead of $L$ a first-order $2\times2$-matrix differential
expression
(i.e., a Dirac-type operator), one arrives at
the ZS hierarchy \cite{ZS72} which is associated with the
nonlinear
Schr\"odinger equation, and more generally, at the AKNS
hierarchy
\cite{AKNS74} which includes the KdV, the mKdV, and the
nonlinear
Schr\"odinger hierarchies as special cases. If one considers
instead
of the second-order differential expression $L$ a differential
expressions of any order, one arrives at the Gelfand-Dickey
hierarchy
\cite{Di91}, \cite{GD76}. Moreover, the formalism is not
confined
to differential
expressions but extends to formal pseudo-differential
expressions
(including rational functions of $d/dx$) needed in connection
with the
sine-Gordon and (modified) Kadomtsev-Petviashvili hierarchies.

\subsection{The Algebra of Commuting Differential Expressions.}
\lb{com}

As seen from \eqref{lax}, stationary (i.e., time-independent)
Lax
equations naturally lead to commuting differential
expressions.
Independently of this fact, the question of commuting
differential
expressions was raised by Floquet \cite{Fl79} in 1879.
Some 25 years
later, the question was again considered by Wallenberg
\cite{Wa03} and
Schur \cite{Sc05}. To this end, Schur developed the algebra
of symbolic,
that is, formal pseudo-differential expressions. He proves,
in
particular, the following statement: if $A$ and $B$ are
two monic
commuting differential expressions then the coefficients
of $A$ are
differential polynomials in the coefficients of $B$ (for a
contemporary
approach to these results, see, e.g., Wilson \cite{Wi79}).
The decisive
step, however, was done by Burchnall and Chaundy in the
1920's. They
proved the following result in \cite{BC23}.

\begin{theorem}
Let $A$ and $B$ be ordinary differential expressions of
relatively prime
orders $m$ and $n,$ respectively. Then $[A,B]=0$ if and
only if
there exists a
polynomial $f$ of the form
\begin{equation} \label{bcpoly}
f(\alpha,\beta)=\alpha^n-\beta^m +
\sum_{\substack{j,k\geq0\\mj+nk<mn}} c_{j,k} \alpha^j \beta^k,
\end{equation}
such that $f(A,B)=0.$
\end{theorem}

$f$ is called the Burchnall-Chaundy polynomial of $A$ and $B$.
In
\cite{BC28} Burchnall and Chaundy constructed differential
expressions
$A$ and $B$ satisfying the equation $f(A,B)=0$ for a given
polynomial
$f$ of the form \eqref{bcpoly} (cf.~also Baker \cite{Ba28},
Burchnall
and Chaundy \cite{BC32}). More recent treatments of the
Burchnall-Chaundy theory can be found, for instance, in
Carlson and
Goodearl \cite{CG80}, Gatto and Greco \cite{GG91}, Giertz,
Kwong, and
Zettl \cite{GKZ81}, Greco and Previato \cite{GP91},
Krichever
\cite{Kr77}, \cite{Kr77a}, Mumford \cite{Mu77},
Previato \cite{Pr96},
and Wilson \cite{Wi85}.

Finally, we briefly return to the Lax pairs $(P,L)$
discussed in
Section~\ref{laxp}. Because of the Burchnall-Chaundy
relationship we
call $L$ (or the set of its coefficients) algebro-geometric
if there
exists a corresponding $P$ such that $[P,L]=0$. We will
provide more
precise definitions in the contexts of the KdV and the
AKNS hierarchies
in Definitions \ref{ag} and \ref{agakns}, respectively.

\subsection{Elliptic Functions in a Nutshell.} \lb{ellfcts}

Elliptic functions provide some of the most important
examples of
algebro-geometric potentials. We present here a very
brief account of
Weierstrass' point of view. For general references see,
for instance,
Akhiezer \cite{Ak90}, Chandrasekharan \cite{Ch85},
Markushevich
\cite{Ma85}, and Whittaker and Watson \cite{WW86}.

A function $f:\bbC\to\bbC\cup\{\infty\}$ with two
periods $a$ and $b$,
the ratio of which is not real, is called doubly
periodic. If all its
periods are of the form $m_1 a+m_2 b$ where $m_1$ and
$m_2$ are integers
then $a$ and $b$ are called fundamental periods of $f$.

A doubly periodic meromorphic function is called elliptic.

It is customary to denote the fundamental periods
of an elliptic
function by $2\omega_1$ and $2\omega_3$ with
$\Im(\omega_3/\omega_1)>0$.
We also introduce $\omega_2=\omega_1+\omega_3$ and
$\omega_4=0$. The
numbers $\omega_1,...,\omega_4$ are called half-periods.
The fundamental
period parallelogram (f.p.p.) $\Delta$ is the half-open
region
consisting of the line segments $[0,2\omega_1)$,
$[0,2\omega_3)$ and the
interior of the parallelogram with vertices $0$, $2\omega_1$,
$2\omega_2$ and $2\omega_3$.

The class of elliptic functions with fundamental periods
$2\omega_1,
2\omega_3$ is closed under addition, subtraction,
multiplication,
division by non-zero divisors and differentiation. If
$f$ is an entire
elliptic function then $f$ is constant. A non-constant
elliptic function
$f$ must have at least one pole in $\Delta$ and the total
number of
poles in $\Delta$ is finite. The total number of poles
of an elliptic
function $f$ in $\Delta$ (counting multiplicities) is
called the order
of $f$. The sum of residues of an elliptic function $f$ at
all its poles
in $\Delta$ equals zero. In particular, the order of a
non-constant
elliptic function $f$ is at least $2$. The total number of
points in
$\Delta$ where the non-constant elliptic function $f$ assumes
the value
$A$ (counting multiplicities), denoted by $n(A)$, is equal
to the order
of $f$. In particular, $n(A)\geq 2$. Furthermore, $s(A)$,
the sum of all
the points in $\Delta$ where the non-constant elliptic
function $f$
assumes the value $A$, is congruent to $s(\infty)$, the sum
of all the
points in $\Delta$ where $f$ has a pole, that is,
$s(A)=s(\infty)+2m_1\omega_1+2m_3\omega_3$, where $m_1$
and $m_3$ are
certain integers.

The function
$$\wp(z;\omega_1,\omega_3)=\frac1{z^2}+
\sum_{\atop{m,n\in\bbZ}
 {(m,n)\neq(0,0)}}
\left(\frac1{(z-2m\omega_1-2n\omega_3)^2} -
 \frac1{(2m\omega_1+2n\omega_3)^2}\right),$$
or $\wp(z)$ for short, was introduced by Weierstrass.
It is an even
elliptic function of order 2 with fundamental periods
$2\omega_1$ and
$2\omega_3$. Its derivative $\wp'$ is an odd elliptic
function of order
3 with fundamental periods $2\omega_1$ and $2\omega_3$.
Every elliptic
function may be written as $R_1(\wp(z))+R_2(\wp(z))\wp'(z)$
where $R_1$
and $R_2$ are rational functions of $\wp$.

The numbers
\begin{align*}
 g_2 &= 60\sum_{\atop{m,n\in\bbZ}{(m,n)\neq(0,0)}}
 \frac1{(2m\omega_1+2n\omega_3)^4}, \\
 g_3 &= 140\sum_{\atop{m,n\in\bbZ}{(m,n)\neq(0,0)}}
 \frac1{(2m\omega_1+2n\omega_3)^6}
\end{align*}
are called the invariants of $\wp$. Since the coefficients
of the
Laurent expansions of $\wp(z)$ and $\wp'(z)$ at $z=0$ are
polynomials of
$g_2$ and $g_3$ with rational coefficients, the function
$\wp(z;\omega_1,\omega_3)$ is also uniquely characterized
by its
invariants $g_2$ and $g_3$. One frequently also uses
the notation
$\wp(z|g_2,g_3)$.

The function $\wp(z)$ satisfies the first order
differential equation
\begin{equation} \label{ps}
\wp'(z)^2 = 4\wp(z)^3-g_2\wp(z)-g_3
\end{equation}
and hence the equations
$$\wp''(z) = 6\wp(z)^2-g_2/2 \text { and } \wp'''(z)=
12\wp'(z)\wp(z)$$
which shows that $-2\wp$ is a stationary solution of the
KdV equation.

The function $\wp'$, being of order $3$, has three zeros
in $\Delta$.
Since $\wp'$ is odd and elliptic it is obvious that these
zeros are the
half-periods $\omega_1,\omega_2=\omega_1+\omega_3$ and
$\omega_3$. Let
$e_j=\wp(\omega_j)$, $j=1,2,3$. Then \eqref{ps} implies that
$4e^3_j-g_2e_j-g_3=0$ for $j=1,2,3$. Therefore
\begin{align*}
 0 &= e_1+e_2+e_3,\\
 g_2 &=-4(e_1e_2+e_1e_3+e_2e_3)=2(e^2_1+e^2_2+e^2_3),\\
 g_3 &= 4e_1e_2e_3 =\frac43 (e^3_1+e^3_2+e^3_3).
\end{align*}

Weierstrass also introduced two other functions denoted by
$\zeta$ and
$\sigma$. The Weierstrass $\zeta$-function is defined by
$$\frac{d}{dz}\zeta(z) = -\wp(z), \quad
 \lim_{z\to0} (\zeta(z)-\frac1z) = 0.$$
It is a meromorphic function with simple poles at
$2m\omega_1+2n\omega_3, m,n\in\bbZ$ having residues $1$.
It is not
periodic but quasi-periodic in the sense that
$$\zeta(z+2\omega_j)=\zeta(z)+2\eta_j, \quad j=1,2,3,4,$$
where $\eta_j=\zeta(\omega_j)$ for $j=1,2,3$ and $\eta_4=0$.

The Weierstrass $\sigma$-function is defined by
$$\frac{\sigma'(z)}{\sigma(z)}=\zeta(z), \quad
 \lim_{z\to0} \frac{\sigma(z)}{z}=1.$$
$\sigma$ is an entire function with simple zeros at the
points
$2m\omega_1+2n\omega_3, m,n\in\bbZ$. Under translation
by a period
$\sigma$ behaves according to
$$\sigma(z+2\omega_j)=-\sigma(z)e^{2\eta_j(z+\omega_j)},
\quad j=1,2,3.$$

Next we recall  the following fundamental theorems.

\begin{theorem} \label{T2.1}
Given an elliptic function $f$ with fundamental periods
$2\omega_1$
and $2\omega_3$, let $b_1,...,b_r$ be the distinct poles of
$f$ in
$\Delta$. Suppose the principal part of the Laurent
expansion near
$b_k$ is given by
$$\sum^{\beta_k}_{j=1} \frac{A_{j,k}}{(z-b_k)^j},
\quad k=1,...,r.$$
Then
$$f(z) = C+\sum^r_{k=1} \sum^{\beta_k}_{j=1}(-1)^{j-1}
 \frac{A_{j,k}}{(j-1)!} \zeta^{(j-1)}(z-b_k),$$
where $C$ is a suitable constant and $\zeta$ is constructed
from
the fundamental periods $2\omega_1$ and $2\omega_3$.
Conversely, every such function is an elliptic function if
$\sum^{r}_{k=1} A_{1,k}=0$.
\end{theorem}

\begin{theorem} \label{T2.2}
Given an elliptic function $f$ of order $n$ with fundamental
periods
$2\omega_1$ and $2\omega_3$, let $a_1,...,a_n$ and
$b_1,...,b_n$ be the
zeros and poles of $f$ in $\Delta$ repeated according to their
multiplicities. Then
$$f(z) = C \frac{\sigma(z-a_1)\cdots\sigma(z-a_n)}
 {\sigma(z-b_1)\cdots\sigma(z-b_{n-1}) \sigma(z-b_n')},$$
where $C$ is a suitable constant, $\sigma$ is constructed from
the fundamental periods $2\omega_1$ and $2\omega_3$ and where
$$b_n'-b_n=(a_1+...+a_n)-(b_1+...+b_n)$$
is a period of $f$.
Conversely, every such function is an elliptic function.
\end{theorem}

Finally, we turn to elliptic functions of the second kind,
the central
object in our analysis. A meromorphic function
$\psi:\bbC\to\bbC\cup\{\infty\}$ for which there exist
two complex constants $\omega_1$ and $\omega_3$ with
non-real ratio and
two complex constants $\rho_1$ and $\rho_3$ such that for
$i=1,3$
$$\psi(z+2\omega_i)=\rho_i \psi(z)$$
is called elliptic of the second kind. We call
$2\omega_1$ and
$2\omega_3$ the quasi-periods of $\psi$. Together with
$2\omega_1$ and
$2\omega_3$, $2m_1\omega_1+2m_3\omega_3$ are also
quasi-periods of
$\psi$ if $m_1$ and $m_3$ are integers. If every quasi-period
of $\psi$
can be written as an integer linear combination of
$2\omega_1$ and
$2\omega_3$ then these are called fundamental quasi-periods.

\begin{theorem} \label{T2.3}
A function $\psi$ which is elliptic of the second kind and
has fundamental
quasi-periods $2\omega_1$ and $2\omega_3$ can always be
put in
the form
$$\psi(z) = C \exp(\lambda z) \frac{\sigma(z-a_1)
\cdots\sigma(z-a_n)}
 {\sigma(z-b_1)\cdots\sigma(z-b_{n})}$$
for suitable constants $C$, $\lambda$, $a_1,...,a_n$ and
$b_1,...,b_n$.
Here $\sigma$ is constructed from the fundamental periods
$2\omega_1$
and $2\omega_3$. Conversely, every such function is elliptic
of the
second kind.
\end{theorem}

\begin{theorem} \label{T2.4}
Given numbers $\alpha_1,...,\alpha_m$ and $\beta_1,...,
\beta_m$ such
that $\beta_k\neq\beta_\ell\,({\rm mod}\,\Delta)$ for
$k\neq\ell$, the
following identity holds
\begin{equation} \label{30071}
\prod_{j=1}^m \f{\sigma(x-\alpha_j)}{\sigma(x-\beta_j)}
 =\sum_{j=1}^m \f{\prod_{k=1}^m \sigma(\beta_j-\alpha_k)}
 {\prod_{\ell=1,\ell\neq j}^m \sigma(\beta_j-\beta_\ell)}
 \f{\sigma(x-\beta_j+\beta-\alpha)}
 {\sigma(x-\beta_j)\sigma(\beta-\alpha)},
\end{equation}
where
$$\alpha=\sum_{j=1}^m \alpha_j \text{ and }\beta=
\sum_{j=1}^m \beta_j$$
and $\sigma$ is constructed from the fundamental periods
$2\omega_1$ and
$2\omega_3$.
\end{theorem}

\begin{proof}[Sketch of proof]
Since this result seems less familiar than those above we
will briefly
sketch its proof. Denote the left and right-hand sides of
\eqref{30071} by $f$ and $g$, respectively. Both $f$ and
$g$ are
elliptic functions of the second kind associated with the
same Floquet
multipliers (with respect to translations by $2\omega_k,$
$k=1,3$).
Their quotient is therefore an elliptic function. Also $f$
and $g$ have
the same poles and zeros taking multiplicities into account.
The
statement about the zeros is a consequence of the identity
(cf.~\cite{WW86}, p.~451)
\begin{equation}
\sum_{j=1}^m \bigg(\f{\prod_{k=1}^m \sigma(\gamma_j-\delta_k)}
{\prod_{\ell=1,\ell\neq j}^m \sigma(\gamma_j-\gamma_\ell)}
\bigg) =0
\text{ if } \sum_{j=1}^m \gamma_j=\sum_{j=1}^m \delta_j.
\end{equation}
Therefore $f/g$ is entire and hence constant. Since all
residues are
equal to one $f=g$.
\end{proof}

\subsection{Hill's Equation and its Spectral Theory.}
\label{hill}

The study of linear homogeneous differential equations with
periodic
coefficients predates the late nineteenth century,
but the
equation $(Ly)(x)=y''(x)+q(x)y(x)=zy(x),$ where $q(x)$ is a
continuous,
real-valued,
periodic function of a real variable $x$ and $z$ is a real
parameter, has generally been called Hill's equation since its
appearance in the study of the
lunar perigee by Hill \cite{Hi77} in 1877. In addition to its
applications in celestial mechanics, this equation
has found countless
applications in quantum mechanics, where it becomes
Schr\"odinger's
equation and is used, for instance, to model crystal
structures of solids.

Periodic differential equations are usually studied by
applying Floquet
theory. Floquet theory (first developed by
Floquet \cite{Fl80},
\cite{Fl83} starting in 1880) specifies the general
structure of
solutions of systems of periodic differential equations.
Consider the
equation ${\ul y}'(x)=Q(x){\ul y}(x),$ where $Q$ is an
$n\times n$
matrix whose entries are continuous (for simplicity) and
periodic with
period $\Omega>0$ and ${\ul y}(x)$ is $\bbC^n$-valued.
Let $\calY$ be
the space of solutions of ${\ul y}'(x)=Q(x){\ul y}(x)$ and
${\calT}_\Omega$ the restriction of ${\ul y}\mapsto {\ul
y}(\cdot+\Omega)$ to $\calY.$ Floquet theory then amounts
to the  study
of the operator ${\calT}_\Omega.$ Since ${\calT}_\Omega$
maps the
$n$-dimensional vector space $\calY$ to itself, the
problem is reduced
to a problem in linear algebra. The eigenvalues and
eigenfunctions of
${\calT}_\Omega$ are called Floquet multipliers and Floquet
functions,
respectively.  For general references on Floquet theory
see, for
instance, Arscott \cite{Ar64}, Coddington and
Levinson \cite{CL85},
Ch.~3, Eastham \cite{Ea73}, Ince \cite{In56}, Sect.~10.8,
Magnus and
Winkler \cite{MW79}, Marchenko \cite{Ma86}, Sect.~3.4,
McKean and van
Moerbeke \cite{MM75}, and Yakubovich and Starzhinskii
\cite{YS75}. In
the special case of Hill's equation $(Ly)(x)=y''(x)+
q(x)y(x)=zy(x),$ we
will denote the translation operator restricted to the
set of solutions
$(y(z,x),y'(z,x))^t\in{\calY}(z)$ of the associated
first-order system,
by ${\calT}_\Omega(z).$

Several differential operators (resp., boundary value
problems)
are studied in connection with Hill's equation,
$$(Ly)(x)=y''(x)+q(x)y(x)=zy(x), \quad q(x+\Omega)=q(x).$$
\begin{itemize}
 \item[(1)] The maximally defined  operator $H$ in
$L^2(\bbR)$
associated
with $L.$
 \item[(2)] Auxiliary operators in $L^2([x_0,x_0+\Omega])$
associated
with
$L$ and certain families of boundary conditions, in
particular,
Dirichlet boundary conditions.
 \item[(3)] The operators in $L^2([x_0,x_0+\Omega])$
associated
with $L$ and
cyclic
boundary conditions $y(x_0+\Omega)=\exp(i\theta)y(x_0),$
$y'(x_0+\Omega)=\exp(i\theta)y'(x_0),$ $\theta\in [0,2\pi),$
in particular, periodic
($\theta=0$)
and anti-periodic ($\theta=\pi$) boundary conditions.
\end{itemize}

Floquet theory provides us with a handle on the problem of
determining
spectral
properties of these operators: the periodic eigenvalues are
given as
the (necessarily real)
zeros of $\tr ({\calT}_\Omega(z))-2,$ while the anti-periodic
eigenvalues are
given as the (necessarily real)
zeros of $\tr ({\calT}_\Omega(z))+2,$ and the spectrum of
$H$ coincides
with the conditional
stability set $\calS (q),$ that is, the set of all values
$z\in\bbR$
such
that $(Ly)(x)=zy(x)$ has a nontrivial bounded solution
with respect to
$x\in\bbR.$ This was first shown by
Wintner \cite{Wi47}, \cite{Wi48} in 1947/48. The conditional
stability set, in turn, may be characterized as the set of all
$z\in\bbR$ such that
$Ly=zy$ has a Floquet multiplier of absolute value one. Since
$$
\det(\calT_\Omega(z))=1,
$$
the Floquet multipliers, being the eigenvalues of
$\calT_\Omega(z),$
 are given as the zeros of
$\rho^2-\rho\tr ({\calT}_\Omega(z))+1$ and we get
$${\mathcal S} (q)=\{z\in\bbR\,|-2\leq
\tr ({\calT}_\Omega(z))\leq 2\}.$$
The conditional stability set consists of countably
(possibly finitely)
many closed intervals (plus possibly a half-line), whose
endpoints
coincide with points where only one (linearly independent)
Floquet
solution exists (which is
necessarily \hbox{(anti-)}perio\-dic). This
was first shown by Hamel \cite{Ha13} in 1913 (see
also Liapunov
\cite{Li99}, who treated the case $y''=\lambda p y$ with
a periodic
function $p$ in 1899, and Haupt \cite{Ha19}, who
corrected a mistake in
Hamel's paper). Hence the \hbox{(anti-)}periodic
eigenvalues determine
the spectrum of $H.$ From the periodic and anti-periodic
eigenvalues
repeated according to their multiplicity and ordered as
a decreasing
sequence denoted by $E_0, E_1,...$ (observing that for
normal operators
the algebraic and geometric multiplicities of eigenvalues
coincide), one
infers
$$\sigma(H)=\bigcup_{j=0}^\infty [E_{2j+1},E_{2j}].$$
Typically, the spectral bands $[E_{2j+1},E_{2j}]$ are
separated by
spectral gaps $(E_{2j+2},\linebreak[0]E_{2j+1})$. However,
since the
\hbox{(anti-)}periodic eigenvalues may be twofold
degenerate, some gaps
(or even all gaps) may close. If only finitely many gaps
are present,
one calls $q,$ the potential coefficient in $L$, a
finite-gap, or a
finite-band potential.

A trivial example of a finite-band potential is of course
the constant
potential $q(x)=c$. The first in this century to discuss
a nontrivial
example was Ince \cite{In40} around 1940, who treated in
depth Lam\'e's
potential
\begin{equation} \lb{lameq}
q(x)=-s(s+1)\wp(x+\omega_3), \quad s\in\bbN,
\end{equation}
with fundamental half periods $\omega_1\in\bbR$ and
$\omega_3\in i\bbR$.
Under these conditions $q$ is real-valued and real-analytic
in $x$.
Ince showed that for all but $2n+1$ values of $z\in\bbR,$
the equation
$y''(x)+q(x)y(x)=zy(x)$ has two linearly independent
\hbox{(anti-)}periodic eigenvalues, that is, $q$ in
\eqref{lameq} is a finite-band
potential (see also Akhiezer \cite{Ak78}, Erdelyi
\cite{Er41}, Turbiner
\cite{Tu89}, and Ward \cite{Wa87}). However, a closer
look at the
classical works of Hermite \cite{He12} and Halphen
\cite{Ha88}
(see also Klein \cite{Kl92}) on
Lam\'e's equation at the end of last century reveals that
these results
were actually well-known (but not yet put in a
Floquet-theoretic
language) as can be inferred, for instance, from
Whittaker-Watson's
treatise \cite{WW86}, Sect.~23.7.

In 1909 Birkhoff \cite{Bi09} compared eigenvalues of
various boundary
value problems associated with $L$ on a finite interval.
His results
show that there is
precisely one Dirichlet eigenvalue in each of the intervals
$[E_{2j+2},E_{2j+1}]$ (i.e., in the closure of the gaps).
In contrast to
the \hbox{(anti-)}periodic eigenvalues, the Dirichlet
eigenvalues
associated with the interval $[x_0,x_0+\Omega]$ vary
with $x_0$ in
$[E_{2j+2},E_{2j+1}]$ if $E_{2j+2}<E_{2j+1}$. Writing
${\bbR}=
\cup_{n\in\bbZ} [x_0+n\Omega,x_0+(n+1)\Omega],$ a
comparison of the
Dirichlet problems on $[x_0,x_0+\Omega]$ and on
$(-\infty,x_0)\cup
(x_0,\infty)$ assuming $q(x)$ to be $\Omega$-periodic,
yields the same
Dirichlet spectra in the spectral gaps
$(E_{2j+2},E_{2j+1})$ with
$E_{2j+2}<E_{2j+1}.$ Explicit formulas for the monotone
rate of change
of various kinds of eigenvalues (including Dirichlet
eigenvalues) with
respect to varying $x_0\in\bbR$ can be found in Kong and
Zettl
\cite{KZ96}, \cite{KZ96a} and the literature therein.
In \cite{GS96}
this phenomenon is related to Green's functions and rank-one
perturbations of resolvents.

If $q$ is a locally integrable, real-valued,
periodic function with period $\Omega>0$ we therefore have
equivalence of the following statements:
\begin{itemize}
 \item[(A)] $q$ is a finite-band potential.
 \item[(B)] For only finitely many values of $z$ the
differential
equation $y''(x)+q(x)y(x)=zy(x)$ fails to have two linearly
independent Floquet
solutions.
 \item[(C)] The Dirichlet boundary value problem on the
interval $[x_0,
x_0+\Omega]$ has only a finite number of eigenvalues
depending on
 $x_0.$
\end{itemize}

Spectral theoretic aspects of complex-valued periodic
potentials $q$
have been investigated by Rofe-Beketov \cite{Ro63} and
Tkachenko
\cite{Tk64} (see also Birnir \cite{Bi86}, \cite{Bi86a},
Kotani
\cite{Ko97}, McGarvey \cite{Mc62}--\cite{Mc65a}, Sansuc
and Tkachenko
\cite{ST96}--\cite{ST97}, and Tka\-chenko \cite{Tk92},
\cite{Tk94},
\cite{Tk96} for recent results). They found that the
spectrum of $H$
and the conditional stability set of $L$ still coincide,
that is,
$${\mathcal S} (q)=\{z\in\bbC\,| -2\leq
\tr ({\calT}_\Omega(z))\leq 2\}.$$
Since the conditional stability set is given as the preimage
of $[-2,2]$ under an entire function, it turns out that
the spectrum of
$H$ consists of countably many (possibly finitely many)
regular analytic
arcs. While the term ``finite-gap potential'' is  now rendered
meaningless, it
still makes perfect sense to call a potential a finite-band
potential if
the spectrum is a finite union of regular analytic arcs.
In contrast to
the real-valued case, these spectral arcs can now cross
each other,
see, for instance, \cite{GW95}, \cite{PT91a} for explicit
examples
exhibiting this phenomenon.

Returning to the real-valued case, the isospectral set
$I(q_0)$ of
a given periodic potential $q_0\in C(\bbR)$ with
period $\Omega>0$
(i.e., the set of all $\Omega$-periodic $q\in C(\bbR)$
whose
$2\Omega$-periodic eigenvalues coincide with that of
$q_0$) turns
out to be a manifold, in fact, a (generally infinite
dimensional) torus
generated by a
product of circles. Each circle is uniquely associated with a
spectral gap $(E_{2j+2},E_{2j+1}),$ $E_{2j+2}<E_{2j+1}$ and
the periodic motion of the Dirichlet eigenvalue in
the gap $(E_{2j+2},E_{2j+1})$ corresponding to the interval
$[x,x+\Omega]$ as a function of
$x\in\bbR.$ In the special case where $q$ is real-valued
and periodic,
and the $L^2$-spectrum of $-d^2/dx^2
+q(x)$ is a half-line $(-\infty,E_0],$ Borg \cite{Bo46}
 proved the
celebrated inverse spectral result $q(x)=E_0$ for all
$x\in\bbR.$ A
quick proof of this uniqueness result and extensions to
real-valued
reflectionless potentials $q$ with associated half-line
spectrum
$(-\infty,E_0]$ follows from the trace formula
proved in \cite{GS96} (observing $\xi(\lambda,x)=1/2$
for $\lambda<E_0$ in the corresponding trace formula (3.1)
for $q(x)$
in \cite{GS96}). Similarly, if $q$ is real-valued and
periodic,
and the $L^2$-spectrum of $-d^2/dx^2+q(x)$ is of the type
$(-\infty,E_2]\cup[E_1,E_0],$ $E_2<E_1,$ Hochstadt
\cite{Ho65} proved
that $q(x)=-2\wp(x+\omega_3+\alpha),$ that is, the
Lam\'e potential
\eqref{lameq} for $n=1,$ where $\wp(x)$ denotes the
Weierstrass elliptic function associated with some period
lattice
$\omega_1>0,$ $-i\omega_3>0$ (depending on $E_0,E_1,$ and
$E_2$) and
$\alpha\in\bbR.$ The isospectral torus
for real-valued periodic potentials with three or more
spectral bands
is described, for instance, in Buys and Finkel \cite{BF84},
Finkel,
Isaacson, and Trubowitz \cite{FIT87}, Gesztesy, Simon,
and Teschl
\cite{GST96}, Gesztesy and Weikard \cite{GW93},
Iwasaki \cite{Iw87},
McKean van Moerbeke \cite{MM75}, and McKean and
Trubowitz \cite{MT76}.

It should be emphasized in this context that the
assumption of
real-valuedness of
$q(x)$ cannot be dropped as shown by the well-known example
$q(x)=\exp(ix),$ with associated $L^2$-spectrum
$(-\infty,0]$ (see
the paragraph following Remark~\ref{rfb} for more details).

\subsection{Periodic KdV Potentials.} \lb{pkdv}

In 1974, Novikov \cite{No74} investigated the Cauchy problem
of the
KdV equation in the case of periodic initial data. He noted that
the right generalization of multi-soliton solutions, which are
stationary solutions of appropriate higher-order KdV equations,
are the
finite-band potentials, that is, he proved the following result.

\begin{theorem}
If $q$ is a real-valued, periodic, stationary solution of an
$n$-th
order KdV equation, then the $L^2(\bbR)$-spectrum
associated with $d^2/dx^2+q(x)$ has at most $n$ finite bands.
\end{theorem}

Within a year, Dubrovin \cite{Du75} and Flaschka \cite{Fl75}
also proved the converse,
and Dubrovin and Novikov \cite{DN75} used these results to
solve the
Cauchy problem of the KdV equation in the case of finite-band
initial data. Hence the statement,
\begin{itemize}
\item[(D)] $q$ is a stationary solution of a higher-order
KdV equation,
\end{itemize}
is equivalent to any of the statements (A)--(C) in
Section~\ref{hill}
for real-valued  $q$.

Moreover, it became clear from these investigations that
generically,
finite-band potentials will only be quasi-periodic with
respect to $x$
and not periodic in $x$.

About the same time, Its and Matveev \cite{IM75}
derived their
celebrated formula for finite-gap solutions $q(x,t)$
in terms of the
Riemann theta function associated with the underlying
hyperelliptic
curve and a fixed homology basis on it
(cf.~\cite{BBEIM94}, Ch.~3,
\cite{GH99}, Ch.~1, \cite{NMPZ84}, Ch.~II). Subsequent
extensions of
this formula to general (matrix-valued) integrable systems
were
developed by Dubrovin \cite{Du77}--\cite{Du83}
and Krichever \cite{Kr77}, \cite{Kr77a}, \cite{Kr83}. A
new approach
to finite-band solutions of the KdV hierarchy in terms
of Kleinian
functions was recently developed by Buchstaber, Enol'skii,
and
Leykin \cite{BEL97}, \cite{BEL98}.

\subsection{Elliptic KdV Potentials.} \lb{ellkdvpot}

While the considerations of the preceding subsections
pertain to general
solutions of the stationary KdV hierarchy, we now
concentrate on the
additional restriction that $q$ be an elliptic function
and hence return
to our main subject, elliptic finite-band potentials
$q$ for
$L=d^2/dx^2+q(x)$, or, equivalently, elliptic solutions
of the
stationary KdV hierarchy. The remarkable finite-gap
example of the
Lam\'e potentials
\eqref{lameq} due to Hermite \cite{He12} and Halphen
\cite{Ha88} in the
last century, brought back into the limelight by
Ince \cite{In40}
around 1940, remained the only
explicit elliptic finite-gap example until the KdV flow
$q_t=\dfrac{1}{4}q_{xxx}+
\dfrac{3}{2}qq_x,$ with initial condition
$q(x,0)=-6\wp(x+\omega_3),$ was explicitly integrated by
Dubrovin and
Novikov \cite{DN75} in 1975 (see also Enol'skii
\cite{En83}--\cite{En84a}, Its and Enol'skii \cite{IE86})
and found to
be of the type
\begin{equation}\label{1.9}
q(x,t) =- 2\sum^3_{j=1} \wp(x-x_j(t))
\end{equation}
for appropriate $\{x_j(t)\}_{1\le j\le 3}$. Due to the
unitary evolution
operator $U_n(t)$ constructed with the help of
$P_{2n+1}(t)$ via
$U_{n,t}(t)=P_{2n+1}(t)U_n(t)$, all potentials $q(x,t)$
in \eqref{1.9}
are isospectral to $q(x,0)=-6\wp(x+\omega_3)$.

In 1977, Airault, McKean and Moser, in their seminal
paper \cite{AMM77},
presented the first systematic
study of the isospectral torus $I_{\bbR}(q_0)$ of
real-valued
smooth
potentials $q_0(x)$ of the type
\begin{equation}\label{1.10}
q_0(x) =-2\sum^M_{j=1} \wp(x-x_j)
\end{equation}
with a finite-gap spectrum. Among a variety of results
they proved that
any element $q$ of $I_{\bbR}(q_0)$ is an elliptic
function of the type
\eqref{1.10} (with different $x_j$), with $M$ constant
throughout
$I_{\bbR}(q_0)$ and $\dim I_{\bbR}(q_0)\le M$. In
particular, if $q_0$
evolves according to any equation of the KdV hierarchy
it remains an
elliptic finite-gap potential. The potential \eqref{1.10}
is intimately
connected with completely integrable many-body systems
of the
Calogero-Moser-type \cite{Ca75}, \cite{Mo75} (see also
Bennequin
\cite{Be93}, Birnir \cite{Bi87}, Calogero \cite{Ca78},
Chudnovsky and
Chudnovsky \cite{CC77}, \cite{Ch79}, Olshanetsky and
Perelomov
\cite{OP81}, and Ruijsenaars \cite{Ru87}). This
connection with
integrable particle systems was subsequently exploited
by Krichever
\cite{Kr80} in his construction of elliptic
algebro-geometric solutions
of the Kadomtsev-Petviashvili equation. In the KdV
context of
\eqref{1.10}, Krichever's approach relies on the
ansatz
\begin{equation} \lb{1.10a}
\psi_0(z,x)=e^{\kappa(z) x}\sum_{j=1}^M A_j(z)
\Phi(x-x_j,\rho(z)),
\end{equation}
for the Floquet solutions of $L_0=d^2/dx^2+q_0(x)$,
where
$$\Phi(x,\rho)=\f{\sigma(x-\rho)}
 {\sigma(x)\sigma(-\rho)}e^{\zeta(\rho) x}$$
(assuming for simplicity the generic case
$x_j\neq x_k \,({\rm
mod}\,\Delta)$ for $j\neq k$).  Applying $L_0$ to
\eqref{1.10a} then
yields an M-sheeted covering of the torus associated
with the
fundamental periods $2\omega_1,2\omega_3$ and hence a
description of the
underlying algebraic curve. (We will briefly comment
on this ansatz in
Remark~\ref{R5.2}.)

The next breakthrough occurred in 1988 when
Verdier \cite{Ve88}
published new explicit examples of elliptic
finite-gap potentials.
Verdier's examples spurred a flurry of activities and
inspired Belokolos
and Enol'skii \cite{BE89}, \cite{BE89a}, Smirnov
\cite{Sm89}, and
subsequently Taimanov \cite{Ta90} and Kostov and
Enol'skii \cite{KE93}
to find further such examples by combining the reduction
process of
Abelian integrals to elliptic integrals (see Babich,
Bobenko and Matveev
\cite{BBM83}, \cite{BBM86}, Belokolos, Bobenko,
Enol'skii, Its, and
Matveev \cite{BBEIM94}, Ch.~7, and Belokolos, Bobenko,
Mateev, and
Enol'skii \cite{BBME86}) with the aforementioned
techniques of Krichever
\cite{Kr80}, \cite{Kr83}. This development finally
culminated in a
series of recent results of Treibich and Verdier
\cite{TV90}--\cite{TV92} where it was shown that a general
complex-valued potential of the form
$$q(x)=-\sum^4_{j=1}d_j \; \wp(x-\omega_j)$$
$(\omega_2 =\omega_1+\omega_3, \; \omega_4=0)$ is a
finite-gap potential
if and only if $d_j/2$ are triangular numbers, that is,
if and only if
$$d_j=s_j(s_j+1) \text{ for some } s_j\in \bbZ,
\; 1\le j\le 4.$$
We shall from now on refer to potentials of the type
$$q(x)=-\sum^4_{j=1} s_j(s_j+1)\wp(x-\omega_j),
 \; s_j\in\bbZ, \; 1\le j\le 4$$
as Treibich-Verdier potentials. The methods of Treibich
and Verdier (see
also Colombo, Pirola, and Previato \cite{CPP94}, Previato
\cite{Pr90},
\cite{Pr94}, Previato and Verdier \cite{PV93}) are based on
hyperelliptic tangent covers of the torus $\bbC/\Lambda$
($\Lambda$
being the period lattice generated by $2\omega_1$
and $2\omega_3$).

The state of the art of elliptic finite-gap solutions up
to 1993 was recently reviewed in a special issue of
Acta Applicandae
Math., see, for instance, Belokolos and Enol'skii
\cite{BE94},
Enol'skii and Kostov \cite{EK94}, Krichever \cite{Kr94},
Smirnov \cite{Sm94}, Taimanov \cite{Ta94}, and
Treibich \cite{Tr94}.
For more recent results see Eilbeck and Enol'skii
\cite{EE94},
\cite{EE94a}, \cite{EE95} and Smirnov \cite{Sm94a},
\cite{Sm96}.
Since a complete characterization of all elliptic
finite-gap solutions
of the stationary KdV hierarchy was still open at that
time, we
developed a new approach to this characterization
problem to be
described in Sections~\ref{agkdvp} and \ref{agaknsp}.
As alluded to
at the end of our introduction, Caloger-Moser-type
models are again
an intensive object of study.

Since we will also discuss stationary rational
solutions of the KdV
hierarchy in Section~\ref{agkdvp} we should mention
the case where
$\wp(x)$ in \eqref{1.10} degenerates into $x^{-2}$ as
discussed, for
instance by Airault, McKean, and Moser \cite{AMM77},
Krichever
\cite{Kr78}, \cite{Kr83a}, Moser \cite{Mo75}, \cite{Mo82},
Pelinovsky
\cite{Pe94}, and Shiota \cite{Sh94}.

\subsection{Linear Differential Equations in the Complex
Domain.}
\lb{lde}

While all the developments described in previous
subsections were
in place around 1993, one final point, the connection
of this subject
to the classical area of differential equations
in the complex domain, was made only around 1994 when
we started to
work on \cite{GW96}. In the following we will remind
the reader about this fundamental, but thus far missing
piece, which
plays a decisive role in the remainder of this review.

In the late 1830's, Lam\'e studied Laplace's equation
in confocal
coordinates. After some appropriate changes of variables
this led to
the differential equation \eqref{lameq}, it's Weierstrass
form. At the
end of the last century, Lam\'e's equation (especially in
Jacobi's
form) was
studied intensively by Hermite \cite{He77}, who obtained
the general
solution of the equation for integer values of $n$ and
any value of $z.$
Picard proceeded to consider first general second-order
equations
\cite{Pi79}, then $n$-th order equations \cite{Pi80} (see
also Floquet
\cite{Fl84}--\cite{Fl84b}, Mittag-Leffler \cite{Mi80},
and Halphen
\cite{Ha84}, \cite{Ha85}, \cite{Ha88}), and finally
first-order systems
\cite{Pi81} whose coefficients are elliptic functions.
Consider the
differential equation
\begin{equation} \label{3.1}
{\ul y}'(x)=Q(x){\ul y}(x),
\end{equation}
where ${\ul y}(x)$ is $\bbC^n$-valued and $Q$ is a
$n\times n$
matrix whose entries are elliptic functions
with a common period lattice spanned by the fundamental
periods
$2\omega_1$ and $2\omega_3$.
Let $\calY$ be the space of solutions of
\eqref{3.1} and denote the restriction of the translation
operator
${\ul y}\mapsto
{\ul y}(\cdot+2\omega_j)$ to $\calY$ by ${\calT}_j.$
Using this
notation, Picard's
theorem reads as follows (see also Akhiezer \cite{Ak90},
Sects.~58, 59,
Burkardt \cite{Bu06}, Ch.~15, Forsyth \cite{Fo59},
Ch.~IX, Gray \cite{Gr86}, Sect.~6.1, Halphen \cite{Ha88},
Ch.~XIII,
Ince \cite{In56}, Sect.~15.6, Krause \cite{Kr97},
Vol.~2, Ch.~3, and Picard \cite{Pi28}, Sect.~III.V).

\begin{theorem} \label{picard}
Assume that the first-order system \eqref{3.1} has a
meromorphic
fundamental system of
solutions. Then there exists at least one solution
$\ul \psi$ which is
elliptic of the second kind, that is, the components
of $\ul \psi$ are
meromorphic and ${\calT}_j{\ul \psi}=\rho_j\ul \psi$
for $j=1,3$ for
suitable constants
$\rho_1,\rho_3\in\bbC\backslash\{0\}.$ If in addition,
one of
the operators
${\calT}_1$ and
${\calT}_3$ has distinct eigenvalues, then there exists a
fundamental system of
solutions of \eqref{3.1} which are elliptic of the
second kind.
\end{theorem}

The explicit Floquet-type structure of solutions
of \eqref{3.1} in terms
of a doubly periodic vector, powers of $x$, powers of
the Weierstrass
zeta-function, and an exponential contribution, has
recently been
determined in \cite{GS98}.

About the time Floquet, Fuchs, Hermite, Mittag-Leffler,
and Picard
(cf.~the historical discussions in Gray \cite{Gr86},
Ch.~VI) developed
the theory of differential equations with elliptic
coefficients, Floquet
\cite{Fl80}, \cite{Fl83} also established his celebrated
results for
linear, homogeneous differential equations with
simply-periodic
coefficients. From a historical perspective it is perhaps
interesting to note
that Floquet assumed the coefficients, as well as the
general
solution of the equation, to be  meromorphic in order to
arrive at the
existence of periodic solutions of the second kind, that
is, he obtained
the precise analog of Picard's result. Only later was it
realized that
his theorem extends to continuous periodic coefficients
on $\bbR$
without any reference to meromorphic fundamental systems.
The solutions
Floquet called ``periodic of the second kind'', are today
generally
called Floquet solutions.

Next, we mention another theorem with a similar flavor
that concerns
differential equations with rational coefficients and
meromorphic
fundamental systems of solutions and hence is applicable
to the study of
rational algebro-geometric KdV potentials. In 1885,
Halphen \cite{Ha85} published the following result.

\begin{theorem} \label{halphen}

Assume that the differential equation $y^{(n)}(x)+
q_1(x)y^{(n-1)}(x)+
\dots +q_n(x)y(x)=0$ has rational coefficients bounded at
infinity and a
meromorphic fundamental system of solutions. Then the general
solution
is of the form $R_1(x)\exp(\lambda_1x)+\dots
+R_n(x)\exp(\lambda_nx),$
where $R_1,\dots, R_n$ are rational functions.
\end{theorem}

It should be emphasized that the principal hypothesis in
Theorems~\ref{picard} and \ref{halphen}, the existence
of a meromorphic
fundamental system of solutions, can be verified in a
straightforward
manner by applying the
Frobenius method (see, e.g., Coddington and
Levinson \cite{CL85},
Sect.~4.8, Forsyth \cite{Fo59}, Ch.~III,
Hille \cite{Hi97}, Ch.~9,
Ince \cite{In56}, Ch.~XVI, and Whittaker and Watson
\cite{WW86}, Ch.~X) to each pole of $q$ in the fundamental
period
parallelogram.

Finally we mention an observation made by Appell \cite{Ap80}.
Let
$y_1(x)$ and $y_2(x)$ be linearly independent solutions of
$y''(x)+u(x)y(x)=0.$ Then
$y_1(x)^2,$ $y_1(x)y_2(x),$ and $y_2(x)^2$ are linearly
independent
solutions of
\begin{equation} \label{appell}
w'''(x)+4u(x)w'(x)+2u'(x)w(x)=0.
\end{equation}
This equation is easily integrated and yields
\begin{equation} \label{appint}
g'(x)^2-2g(x)g''(x)-4u(x)g(x)^2=W(y_1,y_2)^2,
\end{equation}
where $g(x)$ is the product of any two solutions $y_1(x)$
and $y_2(x)$
of
$y''(x)+u(x)y(x)=0$ and $W(y_1,y_2)$ denotes their
($x$-independent)
Wronskian. This innocent
looking fact will be of great importance in our analysis
later.
In fact, a comparison of equations
\eqref{aplax} and \eqref{appell} reveals another connection
between the
KdV hierarchy and the Schr\"odinger-type equation
$y''(x)+u(x)y(x)=0.$
Moreover, since the formal Green's function $G(z,x,,x')$ of
$d^2/dx^2+q(x)$
on the diagonal $x=x'$ is of the type $y_1(z,x)y_2(z,x)
/W(y_1(z),y_2(z)),$ \eqref{appint}, with $u(x)=q(x)-z,$ is
equivalent
to the well-known universal nonlinear second-order differential
equation satisfied by $G(z,x,x)$ (see, also Gelfand and Dickey,
\cite{GD75}, \cite{GD79}).

It should be noted that Drach \cite{Dr18}, \cite{Dr19},
\cite{Dr19a}
(see also \cite{CC84}) used
\eqref{appell} to derive a class of completely integrable
systems now
known as the stationary KdV hierarchy as early as 1918/19.
It appears he
was the first to
make the explicit connection between completely integrable
systems
and spectral theory. More than 55 years later, Gelfand and
Dickey
\cite{GD75}, \cite{GD79} also based some of their celebrated
work
on the KdV hierarchy on \eqref{appell}.

\section{Algebro-Geometric, and Especially, Elliptic KdV
Potentials.}
\lb{agkdvp}

\begin{definition} \label{ag}
Suppose $q$ is meromorphic and let $L$ be the differential
expression
$L=d^2/dx^2+q(x).$ Then $q$ is called an
{\it algebro-geometric KdV
potential} (or simply {\it algebro-geometric}) if $q$ is a
solution of some equation of the stationary KdV hierarchy.
\end{definition}

Equivalently, we could define the meromorphic function
$q$ as
algebro-geometric if any one of the following three
conditions is
satisfied.
\begin{enumerate}
 \item [(1)] There exists an odd-order differential
expression $P$
such that
$[P,L]=0$ (according to the result of Burchnall and Chaundy).
 \item [(2)] There exists an ordinary differential
expression $P$
of odd order and a polynomial $R$ such that $P^2=R(L).$
 \item [(3)] There exists a function $F:\bbC^2\to\bbC_\infty,$
which
is a polynomial in the first variable, meromorphic in the
second, and
which satisfies $F'''(z,x)+4(q(x)-z)F'(z,x)+2q'(x)F(z,x)=0$
(cf.~equation \eqref{aplax}).
\end{enumerate}

It can be shown (see Theorem~6.10 by Segal and Wilson
\cite{SW85})
that any solution of
any of the stationary KdV equations is necessarily
meromorphic. Hence
the assumption that $q$ is meromorphic is actually
redundant in
Definition~\ref{ag}.

\subsection{Periodic KdV Potentials.} \lb{pkdvp}

If $q$ is real-valued, locally integrable, and periodic we
obtain the
equivalence of the following statements from the works of
Birkhoff
and Hamel described in Section~\ref{hill}.
\begin{itemize}
 \item [(i)] $q$ is algebro-geometric with
$P^2=\prod_{j=0}^{2n}(L-\lambda_j).$
 \item [(ii)] $q$ is finite-band with spectrum the union of
$(-\infty,\lambda_{2n}]$ and $n$ compact bands $[\lambda_{2j+1},
\lambda_{2j}],$ $j=0,...,n-1.$
 \item [(iii)] $Ly=zy$ has two linearly independent
Floquet solutions for all $z\in\bbC$ with the exception of
the $2n+1$ values $\lambda_0,...,\lambda_{2n}.$
 \item [(iv)]$q$ has $n$ movable Dirichlet eigenvalues,
precisely one in each of the
closures of the spectral gaps $(\lambda_{2j},\lambda_{2j-1}),$
$j=1,...,n.$
\end{itemize}

As mentioned previously, the classical example for finite-band
potentials are the Lam\'e
potentials $q(x)=-s(s+1)\wp(x+c;\omega_1,\omega_3),$
$n\in\bbN,$ with $c=\omega_3$ purely
imaginary and $\omega_1$ real. However, Lam\'e potentials are
algebro-geometric for general choices of the half-periods
as well as for
general choices of $c\in\bbC.$ This suggests the study of
complex-valued potentials
with inverse square singularities as in \cite{GW96} and
\cite{We98}, and we will subsequently  report on some of
these results.

Let $q$ be a complex-valued, periodic function with period
$\Omega>0,$
which is locally integrable on $\bbR\backslash\Sigma,$ where
$\Sigma\in\bbR$ is a discrete set (i.e., a set without finite
accumulation points). Moreover, $q$ is assumed to be
meromorphic near
each $\xi\in\Sigma$ with principal part
$-s(s+1)/(x-\xi)^2,$ where
$s=s(\xi)\in\bbN.$ Then one can define unique solutions of
initial value
problems of the differential equation $y''(x)+q(x)y(x)=
zy(x)$ on
$\bbR\backslash\Sigma$ (with initial conditions at
$x_0\in\bbR\backslash\Sigma$) by analytic continuation
around the
singularities. Even though the potential is no longer
continuous,
Floquet theory (see Section~\ref{hill}) remains
essentially unchanged.

To apply Floquet theory we first introduce a basis in
${\calY}(z),$
the space
of solutions of $y''(x)+q(x)y(x)=zy(x).$ Let $c(z,\cdot,x_0),
s(z,\cdot,x_0)\in{\calY}(z)$ be defined by the initial
conditions
$c(z,x_0,x_0)=s'(z,x_0,x_0)
=1$ and $c'(z,x_0,x_0)=s(z,x_0,x_0)=0$ (prime denoting the
derivative with
respect to the second variable). Let $\rho_\pm$ denote the
Floquet
multipliers of $y''(x)+q(x)y(x)=zy(x),$ that is, the
eigenvalues
of the translation operator
${\calT}_\Omega(z).$ An important role is played by
$\tr ({\calT}_\Omega(z)),$ which is sometimes called the
Floquet
discriminant and which, in our basis, is
given by
$$\tr ({\calT}_\Omega(z))=c(z,x_0+\Omega,x_0)+
s'(z,x_0+\Omega,x_0).$$
Since the trace is of course independent of the chosen basis
in ${\calY}(z),$ the dependence
of the right-hand side on $x_0$ is only apparent. The
Floquet solutions,
may be expressed as
$$f_\pm(z,x_0,x)= s(z,x_0+\Omega,x_0)c(z,x_0,x)
 +(\rho_\pm-c(z,x_0+\Omega,x_0)) s(z,x_0,x),$$
and their Wronskian is given by
$$W(f_+(z,\cdot,x_0),f_-(z,\cdot,x_0))
 =-s(z,x_0+\Omega,x_0)\sqrt{(\tr ({\calT}_\Omega(z)))^2-4}.$$
Next, consider the function
\begin{equation} \label{03121}
g(z,x)=\frac{f_+(z,x,x_0)f_-(z,x,x_0)}
 {W(f_+(z,\cdot,x_0),f_-(z,\cdot,x_0))}.
\end{equation}
As our notation for $g(z,x)$ suggests, the dependence of this
function on
$x_0$ is only apparent since $f_\pm(z,\cdot,x_1)$ are
just multiples of
$f_\pm(z,\cdot,x_0)$ and the right-hand side of
\eqref{03121} is
independent of normalization. In particular, we may
replace $x_0$ by
$x$ in \eqref{03121} to obtain
$$g(z,x)
 =\frac{-s(z,x+\Omega,x)} {\sqrt{(\tr
({\calT}_\Omega(z)))^2-4}}.$$
The function $s(\cdot,x+\Omega,x)$ is an entire function
of order of
growth $1/2.$ The zeros of $s(\cdot,x+\Omega,x)$ are the
Dirichlet
eigenvalues of $d^2/dx^2+q$ on the interval $[x,x+\Omega]$
and their
order, which we denote by $d(z,x),$ is the algebraic
multiplicity of the
corresponding Dirichlet eigenvalue (we set $d(z,x)=0$ if
$s(z,x+\Omega,x)\neq0$). We also introduce $d_i(z) =
\min\{d(z,x)\,|\,
x\in\bbR\backslash\Sigma\}$ and $d_m(z,x)=d(z,x)-d_i(z),$
the immovable
and movable parts of $d(z,x),$ respectively. The quantity
$\sum_{z\in\bbC} d_m(z,x),$ which is independent of $x,$
is called the
number of movable Dirichlet eigenvalues. Using Hadamard's
factorization
theorem we write $g(z,x)=F(z,x) D(z)$ collecting in $F$ the
factors
depending on $x$ and in $D$ the factors independent of $x.$
Then the
multiplicity of a zero $z$ of $D$ is just $d_i(z),$ while the
multiplicity of a zero $z$ of $F(\cdot,x)$ is $d_m(z,x).$
One then
obtains from Appell's equation
\eqref{appint} that
\begin{equation} \label{09121}
F'(z,x)^2-2F(z,x)F''(z,x)-4(q-z)F(z,x)^2=((\tr
({\calT}_\Omega(z)))^2-4)/D(z)^2.
\end{equation}
Recall that a zero $z$ of $(\tr ({\calT}_\Omega(z)))^2-4$
is an
\hbox{(anti-)}periodic eigenvalue
whose multiplicity we denote by $p(z).$ Since the
left-hand side of
equation \eqref{09121} is entire as a function of $z$ we
obtain the
following result.
\begin{theorem}
There exists an entire function $\ul{R}$ such that
$(\tr({\calT}_\Omega(z)))^2-4=
\ul{R}(z)\linebreak[0]D(z)^2.$ In
particular, $p(z)-2d_i(z)\geq0$ for every $z\in\bbC.$
\end{theorem}
There are, at most, countably many points where $p(z)>0$
since these
points are isolated. Therefore, there are at most countably
many points
where $p(z)-2d_i(z)>0.$ These include all algebraically
simple
\hbox{(anti-)}periodic eigenvalues (where $p(z)=1$) but
may well include
other points too.

We call a Floquet solution $\psi(z_0,\cdot)$ of
$y''(x)+q(x)y(x)
=z_0y(x)$ regular if there exist Floquet solutions
$\psi(z,\cdot)$ of
$y''(x)+q(x)y(x) =zy(x),$ which converge pointwise to
$\psi(z_0,\cdot)$
as $z$ tends to $z_0.$ It was shown in \cite{We98} that
the set of
regular Floquet solutions forms a line bundle on the
topological space
$M_F$ obtained from the curve
$\rho^2-(\tr ({\calT}_\Omega(z)))\rho+1=0$
by desingularization at all points where $p(z)=2d_i(z)>0.$ The
space
$M_F$ can be viewed as a double cover of the complex plane
branched
precisely at all points $z$ where $p(z)-2d_i(z)>0.$ In
particular, the
equation $y''(x)+q(x)y(x)=zy(x)$ has two linearly independent
regular
Floquet solutions if and only if $p(z)-2d_i(z)=0.$ In other
words,
$p(z)-2d_i(z)>0$ indicates a defect in the structure of
regular Floquet
solutions. The number $\df(q)
=\sum_{z\in\bbC}(p(z)-2d_i(z))=\deg(\ul{R}),$ which is a
positive
integer or infinity, will therefore subsequently be called the
Floquet
defect of $q.$ When $q$ is real-valued and nonsingular, then
$p(z)-2d_i(z)$ is always zero, except when $p(z)=1,$ which
forces
$d_i(z)=0.$ In this case $\df(q)$ counts the number of points
$z\in\bbC$
where only one linearly independent Floquet solution exists.
In general,
however, $y''(x)+q(x)y(x)=zy(x)$ may have two linearly
independent
Floquet solutions with only one being regular.

\begin{theorem} \label{t2}

The following statements are equivalent:
\begin{itemize}
 \item [(1)] $\df(q)=2n+1.$
 \item [(2)] There are $n$ movable Dirichlet eigenvalues.
 \item [(3)] There exists a differential expression $P_{2n+1}$
of order
$2n+1$ but none of smaller odd order commuting with $L$. This
differential expression satisfies
$P_{2n+1}^2=R_{2n+1}(z)=\prod_{z\in\bbC}(L-z)^{p(z)-2d_i(z)}$
and hence
$q$ is algebro-geometric.
\end{itemize}
\end{theorem}

\begin{proof}[Sketch of proof]
Generally $p(z)\leq 2$ and $d(z,x_0)\leq1$ for all
suitably large $z.$
Moreover, asymptotically, any Dirichlet eigenvalue is
close to two
\hbox{(anti-)}periodic eigenvalues. Therefore,
$\df(q)=2n+1$ implies
$d_m(z,x_0)$ is different from zero for only finitely many
$z\in\bbC$ and
$\sum_{z\in\bbC}d_m(z,x),$ which equals the degree of
$F(\cdot,x_0),$
must be finite. Equation \eqref{09121} then yields $\deg
F(\cdot,x_0)=n.$ Hence (1) implies (2). The converse of
this follows immediately from equation \eqref{09121}.

Differentiation of equation \eqref{09121} shows that
the third criterion
after Definition~\ref{ag} is satisfied when
$F(\cdot,x_0)$ is a
polynomial. Hence (2) implies (3). We remark here
that $R_{2n+1}$ is a
constant multiple of $\ul{R}$. To prove that (3)
implies (2) one has to
show that the zeros of the function $F_n$ defined by
$P_{2n+1}$ are
precisely the movable Dirichlet eigenvalues of $L.$ This
follows from
applying $P_{2n+1}$ given by
\eqref{09122}, successively to the generalized Dirichlet
eigenfunctions
(i.e., the eigenfunctions corresponding to the algebraic
eigenspace)
associated with the movable Dirichlet eigenvalues of $L$
(cf.~\cite{We98} for more details).
\end{proof}

\begin{remark}
The quantity $p(z)-2d_i(z)$ proves to be of utmost importance.
Determining the points where it takes on positive values,
this quantity
then governs the structure of the line bundle of regular
Floquet
solutions and determines the entire function $\ul{R}(z)
=((\tr({\calT}_\Omega(z)))^2-4)/D(z)^2$, which defines
the algebraic
curve associated with $q$ if $q$ is algebro-geometric and
$\ul{R}$ is a
polynomial. When $q$ is real-valued and locally integrable
on $\bbR,$
then geometric and algebraic multiplicities of Dirichlet and
\hbox{(anti-)}periodic boundary value problems coincide.
Therefore,
$0\leq p(z)\leq2$ and $0\leq d(z,x_0)\leq1.$ Thus, $d_i(z)=1$
enforces
$p(z)=2.$ In addition, if $p(z)=2,$ then all solutions are
\hbox{(anti-)}periodic, implying $d_i(z)=1.$ In this case
the questions,
``When is $p(z)-2d_i(z)>0$?'' and ``When is $p(z)=1$?'', are
equivalent.
Hence, in determining the edges of the spectral gaps, the role
played by
the Dirichlet eigenvalues and, in particular, the distinction
between
movable and immovable Dirichlet eigenvalues, is secondary in
the case of
real-valued locally integrable potentials.
\end{remark}

\begin{remark} \lb{rfb}
One may also show that every algebro-geometric potential is a
finite-band potential, that is, the conditional
stability set (which
coincides with the spectrum of $H$ when no singularities
are present)
consists of finitely many regular analytic arcs.

When $q$ is real and has no singularities, the converse
is also true
(Dubrovin \cite{Du75}). However, in general this is not
the case as
the following example shows. Let $q(x)=\e^{2ix}.$ Then a
fundamental
system of solutions of $y''(x)+q(x)y(x)=-\nu^2y(x)$ is given
in terms of
Bessel functions (cf.~\cite{AS72}, Ch.~9) by
$y_1(x)=J_\nu(i\e^{ix}),$ $y_2(x)=Y_\nu(i\e^{ix}).$ Note
that $y_1$ is
always a Floquet solution with multiplier $\e^{\nu\pi i}.$
Hence
$z=-\nu^2$ is in the conditional stability set if and
only if
$\nu\in\bbR.$ Consequently, ${\mathcal S}(L)=(-\infty,0].$
However,
$s(\nu^2,x_0+\pi,x_0)=\pi J_\nu(i\e^{ix_0})
J_{-\nu}(i\e^{ix_0}),$ which
is entire as a function of $x_0.$ Hence $d_i(-\nu^2)=0$ for
all $\nu\in
\bbC,$ that is, every Dirichlet eigenvalue is movable. Thus,
$\df(q)=\infty,$ and hence $q$ is not algebro-geometric. More
general examples of this type have been studied
systematically by
Gasymov \cite{Ga80}, \cite{Ga80a}, Guillemin and
Uribe \cite{GU83},
and Pastur and Tkachenko \cite{PT88}, \cite{PT91}.
\end{remark}

\subsection{Picard-KdV Potentials.} \lb{pickdvp}
\begin{definition} \label{defPicard}
Let $q$ be an elliptic function. Then $q$ is called a {
\it Picard-KdV
potential} (or simply a {\it Picard potential})
if the equation $y''(x)+q(x)y(x)=zy(x)$ has a meromorphic
fundamental system of solutions with respect to $x$ for all
values of
the spectral parameter $z\in\bbC.$
\end{definition}

\begin{theorem} \label{T5.1}
If $q$ is a Picard potential then it may be represented as
\begin{equation} \label{5.10}
q(x)=C-\sum_{j=1}^m s_j(s_j+1) \wp(x-b_j)
\end{equation}
for suitable integers $s_1,...,s_m$ and complex numbers
$C,$ $b_1,
...,b_m,$ where the $b_j$ are pairwise distinct
${\rm mod}\,\Delta.$
\end{theorem}

\begin{proof}[Sketch of proof]
Every singularity of $y''(x)+q(x)y(x)=zy(x)$ must be a regular
singular
point with integer indices. From the partial fraction
expansion for
elliptic functions (Theorem \ref{T2.1}) one obtains
$$q(x)=C-\sum_{j=1}^m (s_j(s_j+1) \wp(x-b_j)+
B_j\zeta(x-b_j)).$$
In a vicinity of $b_j$ there is a solution of the form
$$\psi(x)=(x-b_j)^{-s_j} \sum_{k=0}^\infty
\beta_k (x-b_j)^k,$$
where $\beta_0=1.$ Next we use the Frobenius method to show
that $B_j=0.$ Let
$$q(x)=\frac{-s_j(s_j+1)}{(x-b_j)^2}+\frac{B_j}{x-b_j}
 +\sum_{k=0}^\infty C_{j,k}(x-b_j)^k$$
and insert $\psi$ into the differential equation
$y''+(q-z)y=0$ to get
\begin{align*}
0=&f(-s_j)(x-b_j)^{-s_j} + \{f(1-s_j)\beta_1+G_1\}
(x-b_j)^{1-s_j}\\
&+\cdots+\{f(k-s_j)\beta_k+G_k\} (z-b_j)^{k-s_j}+\cdots,
\end{align*}
where $f(\ell)=(\ell+s_j)(\ell-s_j-1)$ and
$$G_k=B_j \beta_{k-1} + (C_{j,0}-z)\beta_{k-2}
 + C_{j,1}\beta_{k-3} + \cdots + C_{j,k-2} \beta_0.$$
Now $f(-s_j)=0$ and $\beta_1,...,\beta_{2s_j}$ may be
determined
recursively so that the coefficients of $(x-b_j)^{1-s_j},...,
(x-b_j)^{s_j}$ vanish. But since $f(s_j+1)=0,$ the
coefficient of
$(x-b_j)^{s_j+1}$ is just $G_{2s_j+1},$ which therefore must
vanish for
all $z\in\bbC.$ On the other hand, if $B_j\neq0,$ one can
show by
induction that $G_{2s_j+1}$ is a polynomial in $z$ of degree
$s_j.$ This
contradiction completes the proof.
\end{proof}

If $q$ is a Picard potential then, by Picard's theorem
(Theorem~\ref{picard}), the equation $y''(x)+q(x)y(x)=
zy(x)$ has at
least one solution which is elliptic of the second kind.
Using Theorem
\ref{T2.3} and the special structure of the Picard potential
\eqref{5.10} this solution may be represented as
\begin{equation} \label{psia}
\psi_{a(z)}(x)=\frac{\prod_{j=1}^s \sigma(x-a_j(z))}
 {\prod_{j=1}^m \sigma(x-b_j)^{s_j}}\exp(\lambda_{a(z)} x),
\end{equation}
where $s=\sum_{j=1}^{m} s_j$ and $a(z)=(a_1(z),\dots,a_s(z)).$
At $b_j$ the function $\psi_a$ has a pole
of order $s_j$ or a zero of order $s_j+1.$ For later notational
purposes we allow for $s_j=0,$ in which case $\psi_a$ has
either no pole
and no zero or a simple zero. For subsequent use we define
\begin{align*}
 M_1 &= \{j\in\{1,...,m\}\,|\, \ord_{b_j}(\psi_a)=-s_j\},\\
 M_2 &= \{j\in\{1,...,m\}\,|\, \ord_{b_j}(\psi_a)=s_j+1\}.
\end{align*}
The function $\psi_a$ is a solution of $y''(x)+q(x)y(x)=
zy(x)$ if
and only if
\begin{align}
&\lambda_a+\sum_{j=1}^s \zeta(b_r-a_j)-
\sum_{\substack{j=1\\j\neq r}}^m
 s_j \zeta(b_r-b_j)=0, \label{lambda}\\
 &z=C-\sum_{j=1}^s (1-2s_r)\wp(b_r-a_j)
 -\sum_{\substack{\scriptstyle j=1\\j \neq r}}^m
s_j(s_j+2s_r)\wp(b_r-b_j), \label{energy}
\end{align}
where $r$ is chosen such that $s_r\neq0$ and
$\ord_{b_r}(\psi_a)=-s_r$
which is always possible. In the case of a Lam\'e potential
these
conditions are recorded, for instance, by Burkhardt \cite{Bu06}.
The subscript $a$ in
$\lambda_a$ expresses the dependence of $\lambda_a$ on
$a=(a_1,...,a_s),$ which in turn depends on $z.$

In \cite{GW95c} we have developed a method for even Picard
potentials (i.e., potentials $q$ satisfying
$q(x_0+x))=q(x_0-x)$ for some $x_0\in\bbC$) to determine all
points where two
regular Floquet solutions fail to exist. (For
simplicity we will assume $x_0=0$ from now on.) First
note that
an even Picard
potential is of the form
\begin{equation}
q(x)=C-\sum_{k=1}^4 s_{k}(s_{k}+1)\wp(x-\omega_k)
 -\sum_{k=1}^{\tilde m} r_k(r_k+1)[\wp(x-b_k)+
\wp(x+b_k)], \lb{ep}
\end{equation}
where the $b_k$ are pairwise distinct and different from
half-periods,
the $s_k$ are nonnegative and the $r_k$ are positive.
If $\tilde m=0,$
that is, if
$$q(x)=C-\sum_{k=1}^4 s_{k}(s_{k}+1) \wp(x-\omega_k),$$
then $q$ is called a Treibich-Verdier potential
following the work of
Verdier \cite{Ve88} and Treibich and Verdier
\cite{Tr89}--\cite{TV97}.
If in addition, only one of the numbers $s_k$ is
different from zero,
then $q$ is a Lam\'e potential. In order to make use
of previous results
we will adopt the following notation: $m=2\tilde m+4,$
$b_{k+\tilde
m}=-b_k,$ $s_{k+\tilde m}=r_k$ for $k=1,...,\tilde m,$
and $b_{k+2\tilde
m}=\omega_k,$ $s_{k+2\tilde m}=s_k$ for $k=1,...,4.$ Let
$\hat s$ be the
number of $a_j$'s which do not appear in $\{b_1,...,b_m\}.$

If $q$ is an even Picard potential of the type \eqref{ep},
and if $\psi_a$ given by \eqref{psia}
is a solution of the differential equation
$y''(x)+q(x)y(x)=zy(x),$
then so is the
function $\psi_{-a},$ which is obtained by replacing
every $a_j$ with
$-a_j$ in \eqref{psia} and \eqref{lambda}, since
$\psi_{-a}(x)
=(-1)^{s_1+s_2+s_3}\psi_a(-x).$

Next we compute the Wronskian of the two solutions
$\psi_a$ and
$\psi_{-a}.$ One obtains an expression which involves
$x,$ but since the
Wronskian does not depend on $x,$ one may evaluate it
anywhere. Choosing
$x=a_\ell$ for any $a_\ell$ which does not appear in
$\{b_1,...,b_m\},$
one finds that the Wronskian is a nonzero multiple of
$$\frac{\sigma(2a_\ell)}{\sigma(a_{\ell}-b_r)
\sigma(a_{\ell}+b_r)}
 \prod_{\substack{j=1\\j\neq \ell}}^{\hat s}
 \sigma(a_{\ell}-a_j)\sigma(a_{\ell}+a_j).$$
Since $a_\ell$ is different from all the $b_j$ and, in
particular, different from the half-periods, one infers
$\sigma(2a_\ell)\neq0.$ Also $a_\ell \neq a_k$ if $k\neq
\ell.$ Therefore we find that the Wronskian is zero if
and only if
$\sigma(a_{\ell}+a_j)=0$ for some $j\in\{1,...,\ell-1,
\ell+1,...,\hat
s\}$ and hence $a_j=-a_\ell\,({\rm mod}\,\Delta$). In
particular, we
find that
the number $\hat s$ is even and define $d=\hat s/2.$

Choosing now $x=b_\ell$ for any $\ell\in M_2$ the Wronskian
can be
written as a nonzero multiple of
$$\frac{\sigma(2b_\ell)^{2s_\ell+1}}
{\sigma(b_{\ell}-b_r)\sigma(b_{\ell}+b_r)}
\frac{\prod_{\substack{j\in M_2\\j\neq \ell}}
{\sigma(b_{\ell}-b_j)^{2s_j+1}\sigma(b_{\ell}+b_j)^{2s_j+1}}}
{\prod_{\substack{j=1\\j\neq \ell}}^{m}
\sigma(b_{\ell}-b_j)^{2s_j}},$$
which is zero if and only if $b_\ell$ is a half-period or
if there is a
$j\in M_2$ such that $b_j=-b_\ell \,({\rm mod}\,\Delta$).

In summary we have found the following: if $\psi_a$ and
$\psi_{-a}$ are
linearly dependent solutions of $\psi''+q\psi=z\psi,$ then
some of the
numbers $a_1,$ ..., $a_s$ may be half-periods while all
others appear in
pairs $(a_j,a_{\ell_j})$ with $a_{\ell_j}=-a_j.$ Moreover, if
$a_j$ is
equal to a half-period $\omega_k,$ which is a pole of $q$
of the form
$-r_k(r_k+1)/(z-\omega_k)^2,$ then there are exactly
$2r_k+1$ of the
$a_{\ell}$ which are equal to this half-period. If $a_j$
equals a
pole $b_{\ell}$ of the form $-s_{\ell}(s_{\ell}+
1)/(z-b_{\ell})^2,$ where
$b_{\ell}$ is not a half-period, then there are exactly
$2s_{\ell}+1$ of
the $a_m$ which are equal to this pole and exactly
$2s_{\ell}+1$ other
$a_m$'s which are equal to the pole $-b_{\ell}.$

This information is now being used to rewrite the solution
$\psi_a$ of
$\psi''+q\psi=z\psi$ for those values of the spectral
parameter $z$
where $W(\psi_a,\psi_{-a})=0$ as a product of two functions.
The first
function is fixed, depending only on the poles of the
potential $q,$ on
the half-periods, and the exponents associated with these.
The second
function is a polynomial in $\wp(x),$ whose coefficients
depend on those
$a_j$ which are neither half-periods nor poles of $q$ and
which are as
yet undetermined. According to the above argument there must
be an even
number, $2d,$ of those, and half of them are just negatives
of the other
half.

Hence we define $t_\ell=\ord_{b_\ell}(\psi_a)$ for
$\ell=1,...,2\tilde m+4$ and obtain $\psi_a(x)=
f(x) Q(\wp(x)),$ where
\begin{align*}
f(x) &= \e^{\lambda_a x}
\left(\prod_{k=1}^4 \sigma(x-\omega_k)^{t_{k+2\tilde m}}\right)
\left(\prod_{\ell=1}^{2\tilde m}
\sigma(x-b_{\ell})^{t_{\ell}}\right)
\sigma(x)^{2d},\\
Q(\wp(x))&=\prod_{j=1}^d(\wp(x)-\wp(a_j))=\sum_{j=0}^d c_j
\wp(x)^j.
\end{align*}
Here we used the fact that $\sigma(z-a_j)\sigma(z+a_j)
=-\sigma(z)^2\sigma(a_j)^2 (\wp(z)-\wp(a_j)).$ Moreover, we
dropped the
non-zero constant factor $(-1)^d\prod_{j=1}^d \sigma(a_j)^2.$
This
yields
\begin{align*}
&\psi_a''(x)+q(x)\psi_a(x) \\
 =&f(x) \left\{ \sum_{k=0}^{d} \sum_{j=0}^d S_{k+1,j+1}
c_j \wp(x)^k
 + \frac{\sum_{k=1}^{\tilde m} \sum_{j=0}^d T_{k,j+1} c_j
\wp(x)^{k-1}}{\prod_{j=1}^{\tilde m} (\wp(z)-\wp(b_j))}
\right\}
\end{align*}
for suitable constants $S_{j,k}$ and $T_{j,k}$ depending
on $q$
and the numbers $t_\ell.$ Let $S$ and $T$ be the matrices
with entries $S_{j,k}$ and $T_{j,k}.$

Hence we obtain a solution $\psi_a$ of the equation
$y''(x)+q(x)y(x)
=zy(x)$
satisfying $W(\psi_a,\psi_{-a})=0$  if and only if
$$S\gamma=z\gamma \text{ and } T\gamma=0,$$
where $\gamma=(c_0,...,c_d)^T.$

For any given even Picard potential there are several
(but finitely
many) choices to distribute some (or all) of the parameters
$a_1,...,a_s$ among the half-periods and/or poles of
$q.$ Accordingly,
there are several (but finitely many) of the above
described constraint
eigenvalue problems to solve in order to find all the
values of the
spectral parameter $z\in\bbC$ where
$W(\psi_a,\psi_{-a})=0.$ In each
case there
are only finitely many eigenvalues of the associated
matrix $S,$ some
(or possibly all) of which may be in contradiction to
the constraint
$T\gamma=0.$ Thus, we proved the following result.
\begin{theorem} \label{T5.4a}
Let $q$ be an even Picard potential. Then, for every
complex number $z,$
the differential equation $y''(x)+q(x)y(x)=zy(x)$ has a
solution
$\psi_a$ of the
form \eqref{psia}, that is, a solution which is elliptic of
the second
kind. Similarly, the function $\psi_{-a}$ is a solution of
the same
equation {\rm (}for the same value of $z${\rm )} and also
elliptic of the second kind.
With respect to any period of $q,$ the functions $\psi_a$
and $\psi_{-a}$
are regular Floquet solutions. For all but a finite number
of values of
$z\in \bbC$ these two solutions are linearly independent
and therefore
$p(z)-2d_i(z)=0.$ Hence $\df(q)<\infty,$ that is, $q$ is
algebro-geometric.
\end{theorem}

In the case of a Treibich-Verdier potential
\begin{equation}
q(x)=-\sum_{j=1}^4 s_j(s_j+1) \wp(x-\omega_j),\quad
s_j\in\bbN\cup\{0\}, \,\, j=1,\dots,4,
\end{equation}
the matrix $T$ is absent and therefore one only has to
find the
eigenvalues of $S$ for all possible choices of the numbers
$t_\ell.$
This was performed in \cite{GW95a} for Lam\'e potentials
and in
\cite{GW95b} for Treibich-Verdier potentials. The arithmetic
genus of
the curve $P^2=R(L)$ associated with $q$ is given in the
following
table, where $s=s_1+s_2+s_3+s_4$ (and, without loss of
generality, $s_1
\geq s_2 \geq s_3 \geq s_4 \geq 0$).

\vspace*{.4cm}

\begin{center}
\begin{tabular}{|c|c|c|c|}
\hline $s$&&$\hbox{\rm \# of finite branch points}$&
\hbox{\rm genus}\\ \hline
even & $s_2+s_3\leq s_1+s_4$ &    $2s_1+1$ & $s_1$\\ \hline
even & $s_2+s_3\geq s_1+s_4$ &    $s_1+s_2+s_3-s_4+1$ &
$\f{s}{2} -s_4$\\
\hline
odd  & $s_2+s_3+s_4<s_1$     &    $2s_1+1$ & $s_1$ \\ \hline
odd  & $s_2+s_3+s_4>s_1$     &    $s_1+s_2+s_3+s_4+2$ &
$\f{s+1}{2}$\\
\hline
\end{tabular}
\end{center}

\vspace*{.4cm}

Shortly after this table appeared in our paper \cite{GW95b},
Armando
Treibich kindly sent us the following formula for the
arithmetic genus
of the Treibich-Verdier curve. It is given by
$$\max\{2s_1,s+1-(1+(-1)^s)(s_4+(1/2))\}/2$$
in agreement with our column on the right.

\begin{remark} \lb{R5.2}
We briefly return to the Picard potential $q(x)$ in
\eqref{5.10}.
Assuming first the special case where $s_j=1,$
$j=1,\dots,m,$ that is,
$$q(x)=-2\sum_{j=1}^m \wp(x-b_j), \quad b_k\neq b_\ell \,
 (\text{mod}\,\Delta) \text{ for } k\neq \ell,$$
Theorem \ref{T2.4} yields Krichever's ansatz \eqref{1.10a} in
\cite{Kr80}, that is,
\begin{equation}
\psi_{a(z)}(x)= \exp(\lambda_{a(z)} x) \prod_{j=1}^m
\f{\sigma(x-a_j(z))}{\sigma(x-b_j)} 
=e^{\kappa(z) x}\sum_{j=1}^m A_j(z)\Phi(x-b_j,\rho(z)),
\label{kr}
\end{equation}
where
$$\Phi(x,\rho)
 =e^{\zeta(\rho)x}\sigma(x-\rho)\sigma(x)^{-1}
\sigma(-\rho)^{-1}$$
and where $\kappa$, $A_j$ and $\rho_j$ are suitably chosen.

In the general case, where
\begin{equation}
q(x)=-\sum_{j=1}^m s_j(s_j+1)\wp(x-b_j),
\quad b_k\neq b_\ell \,
(\text{mod}\,\Delta) \text{ for } k\neq \ell,
\end{equation}
one can use an extension of Theorem \ref{T2.4} (using, e.g.,
l'Hospital's rule as several $b_j$'s are confluenting into
a $b_{j_0}$
and converting $b$-derivatives into an $x$-derivative)
to arrive at the
corresponding analog
\begin{align} \lb{kri}
\psi_{a(z)}(x)&= \exp(\lambda_{a(z)} x) \f{\prod_{j=1}^s
\sigma(x-a_j(z))} {\prod_{j=1}^m \sigma(x-b_j)^{s_j}}  \\
&=e^{\kappa(z) x}\sum_{j=1}^m \sum_{k=0}^{s_j-1} A_{j,k}(z)
\Phi^{(k)}(x-b_j,\rho(z)), \quad s=\sum_{j=1}^m s_j \no
\end{align}
of \eqref{kr}. The extended ansatz \eqref{kri} was
recently used by
Enol'skii and Eilbeck \cite{EE94}, \cite{EE95}, Enol'skii
and Kostov
\cite{EK94}, \cite{KE93}, in the context of Treibich-Verdier
potentials.
The special case $m=1,$ $s_1=2$ can be found in
Hermite \cite{He12},
p.~374--377 (see also Forsyth \cite{Fo59}, p.~475--476),
the general
case \eqref{kri} is discussed in Krause's monograph
\cite{Kr97}, Vol.~1,
p.~292--296, Vol.~2, p.~183, 259--264.

We emphasize that Picard's theorem, Theorem~\ref{picard},
yields the
$\sigma$-function representations of $\psi_{a(z)}(x)$
in \eqref{kr} and
\eqref{kri} and hence Krichever's ansatz \eqref{1.10a} and
its extension
\eqref{kri}. In particular, an alternative characterization
of all
stationary elliptic KdV and AKNS potentials to the one
provided in
Theorem~\ref{t3.12} and Theorem~\ref{t5.4}, respectively, can
be based
on these observations (cf.~\cite{GW99}). In the remainder of
this review
we do not pursue this avenue but stress Floquet-theoretic and
Green's
function methods instead.

\end{remark}
\subsection{Necessary Conditions for a KdV Potential to be
Algebro-Geometric.} \lb{nag}

\begin{theorem} \label{mt1}
Suppose $q$ is an algebro-geometric KdV potential. Then
$(Ly)(x)=y''(x)+q(x)y(x)=zy(x)$ has a meromorphic
fundamental system of
solutions with respect to $x$ for all values of the spectral
parameter
$z\in\bbC.$
\end{theorem}

\begin{proof}[Sketch of proof]
First one shows that every pole of $q$ is of second
order. This
follows since only in this case there is a balance
between the growth of
the order of the poles of $f_j'''$ and $4qf_j'+2q'f_j$
considering the recursion relation \eqref{recrel}. Such
a balance is
necessary since eventually all $f_j$ vanish. Next one
shows by a similar
argument that the leading term of the Laurent expansion
of $q$ about a
pole $x_0$ must be $-s(s+1)/(x-x_0)^2$ for some integer
$s.$ This
guarantees that the exponents of the singularity $x_0$ of the
differential equation $y''(x)+q(x)y(x)=zy(x)$ are integers.
Finally
one has to show
the absence of logarithmic terms in the solutions of
$y''(x)+q(x)y(x)
=zy(x)$ which
follows from a careful analysis using the Frobenius method.
More details
are provided in \cite{We98a}.
\end{proof}

We note that $\tau$-function results in Segal and Wilson
\cite{SW85}
imply Theorem~\ref{mt1}. (In the case of nonsingular
curves $\calK_n$
this simply follows from the standard theta-function
representation
of the Baker-Akhiezer function.) While Segal and Wilson
rely on
loop group techniques (and study the Gelfand-Dickey
hierarchy), the
above proof represents a completely elementary alternative
for the
KdV hierarchy.

\subsection{Sufficient Conditions for a KdV Potential to be
Algebro-Geometric -- Characterizations of Elliptic, Simply
Periodic,
and Rational Stationary KdV Solutions.} \lb{sag}

\begin{theorem} \label{mt2}
$q$ is an algebro-geometric potential if, for each
$z\in\bbC,$
$y''(x)+q(x)y(x)=zy(x)$ has a meromorphic fundamental system
of solutions with respect to $x$ and if one of
the following three conditions is satisfied.
\begin{itemize}
 \item [(i)] $q$ is rational and bounded near infinity.
 \item [(ii)] $q$ is simply periodic with period
$\Omega$ and there exists a
 positive number $R$ such that $q$ is bounded in
$\{x\in\bbC \,|\,|\Im(x/\Omega)|
 \geq R\}.$
 \item [(iii)] $q$ is elliptic.
\end{itemize}
\end{theorem}

\begin{proof}[Sketch of proof]
We will first sketch the proof in \cite{GW96} which
treats the case
where $q$ is elliptic and prominently uses the double
periodicity
of $q.$

Let $q$ be an elliptic function with fundamental periods
$2\omega_1$ and
$2\omega_3.$ Assume, without loss of generality, that
$\Im(\omega_3/\omega_1)>0.$ Introduce
$t_j=\omega_j/|\omega_j|$ and
define
$$q_j(x)=t^2_j q (t_jx).$$
The transformation $\xi=t_jx,$ $\psi(\xi) = w(x)$ transforms
$\psi''+q\psi=z\psi$ into $w''+q_jw=t_j^2zw.$

Note that $q_j$ has the period $2|\omega_j|.$ With the
aid of Rouch\'e's
theorem, one may determine the asymptotic distribution of the
\hbox{(anti-)}periodic eigenvalues of $d^2/dx^2+q_j$ and
verify that
they all stay close to the real axis while their real parts
tend to
$-\infty.$ (Never mind that $q_j$ may have inverse square
singularities,
see \cite{We98}.) Equivalently, all
$2\omega_j$-\hbox{(anti-)}periodic
eigenvalues of $d^2/dx^2+q$ lie in the half-strip $\Sigma_j$
given by
$$\Sigma_j=\{z\in\bbC\,|\, |\Im(t^2_jz)|\leq C, \;
 \Re(t^2_jz) \leq C\}, \quad j=1,3$$
for some constant $C>0.$ Note that the angle between
the strips
$\Sigma_1$ and $\Sigma_3$ is positive and less than $2\pi$ and
therefore, they intersect only in a finite part of the
complex plane.
Hence, for every point $z$ outside a sufficiently large compact
disk, at least one of the
translation operators ${\calT}_1$ or ${\calT}_3$ has distinct
eigenvalues. Assuming,
according to our hypothesis, that all solutions of
$y''(x)+q(x)y(x)
=zy(x)$ are
meromorphic, we may invoke Picard's theorem to conclude that
there
are two linearly independent solutions which are elliptic of
the second
kind for any $z$ outside that disk. These solutions are Floquet
solutions with respect to any period
of $q$ and since the disk is compact, we have shown that there
are at most finitely many
points which lack two linearly independent Floquet solutions.
Applying Theorem~\ref{t2} then proves that $q$ is
algebro-geometric.

Next we sketch the proof for the case when $q$ is rational.
That proof
can easily be extended for the simply periodic and elliptic
potentials.
In the latter case one uses the algebraic properties of elliptic
functions rather than their double periodicity. For more details
see
\cite{We98a}.

Suppose that $q$ is rational and bounded at infinity. Let $z_0
=\lim_{x\to\infty} q(x).$ From Halphen's theorem
(Theorem~\ref{halphen})
we obtain for $z\neq z_0$ the existence of linearly independent
solutions
$$y_\pm(z,x)=R_\pm(z,x)\exp(\pm\sqrt{z-z_0}x),$$
where $R_\pm(z,\cdot)$ are rational functions. The poles of
$R_\pm(z,\cdot)$ are determined as the singular points of the
differential equation and their orders as the exponents of the
corresponding singularities. Next define the function $g(z,x)
=y_+(z,x)y_-(z,x).$ Then there exists a polynomial $v$ in $x$
such that
$$v(x)^2 g(z,x)=\sum_{j=0}^d c_j(z) x^j.$$
Next note that the functions $v^2 g(z,x),$ $v^3 g'(z,x),$ $v^4
g''(z,x),$ and $v^5 g'''(z,x)$ are polynomials in $x$ whose
coefficients
are homogeneous polynomials of degree one in $c_0,...,c_d.$
 Also $v^2q$
and $v^3q'$ are polynomials. Hence $v^5(g'''+4(q-z)g'+2q'g)$
is also a
polynomial in $x,$ whose coefficients are homogeneous
polynomials of
degree one in $c_0,\dots,c_d.$ The coefficients of the
polynomials
$c_\ell$ in this last expression are polynomials in $z$ of
degree at
most one, that is,
\begin{equation} \label{08011}
v^5(g'''+4(q-z)g'+2q'g)=\sum_{j=0}^N
\sum_{\ell=0}^d (\alpha_{j,\ell}
+\beta_{j,\ell}z) c_\ell x^j
\end{equation}
for suitable numbers $N,$ $\alpha_{j,\ell},$ and
$\beta_{j,\ell}$ which
only depend on $q.$ From Appell's equation \eqref{appell}
it follows
upon differentiation that the expression \eqref{08011} vanishes
identically. This gives rise to a homogeneous system of
$N+1$ linear
equations which has a nontrivial solution. Solving
the system shows now that the coefficients $c_\ell$ are
rational
functions of $z.$ Therefore
$$g(z,x)=\frac{F(z,x)}{\gamma(z)},$$
where $F(\cdot,x)$ and $\gamma$ are polynomials and
$F(z,\cdot)$ is a
rational function. Hence $q$ is algebro-geometric by item (3)
following Definition~\ref{ag}.
\end{proof}

Combining the results of this subsection and its preceding
one yields
the following explicit characterization of all elliptic
algebro-geometric KdV potentials (a problem posed, for
instance, by
Novikov, Manakov, Pitaevskii, and Zakharov \cite{NMPZ84},
p.~152),
originally proven in \cite{GW96} (see also \cite{GW95d},
\cite{GW98a}).

\begin{theorem} \label{t3.12}
Let $q$ be an elliptic function. Then $q$ is an elliptic
algebro-geo\-metric KdV potential if and only if it is a Picard
potential.
\end{theorem}

Similarly, these results characterize the stationary rational
KdV
potentials vanishing at infinity studied, for instance, by
Adler and
Moser \cite{AM78}, Airault, McKean, and Moser \cite{AMM77},
Calogero
\cite{Ca75}, \cite{Ca78}, Chudnovsky and Chudnovsky \cite{CC77},
\cite{Ch79}, Grinevich \cite{Gr82}, Krichever \cite{Kr78},
Matveev
\cite{Ma79}, and Sokolov \cite{So78}.


\section{Algebro-Geometric and Especially, Elliptic AKNS
Potentials.}
\lb{agaknsp}

Most of the results in this section are taken from \cite{GW98}.
Since we
take here a slightly different point of departure, the notation
differs
a bit from that in \cite{GW98}.

\subsection{The AKNS Hierarchy.} \lb{akns}

Let $L=Jd/dx+Q(x),$ where
\begin{equation} \label{18021}
J=\begin{pmatrix}i&0\\0&-i\end{pmatrix} \text{ and }
Q(x)=\begin{pmatrix}0&-iq(x)\\ip(x)&0\end{pmatrix}.
\end{equation}
Note that $J^2=-I,$ the $2\times 2$ identity matrix, and
that $JQ+QJ=0.$
As mentioned in Section~\ref{laxp} one may consider the Lax
pair $(P_\s
,L)$, where $P_\s $ is a $2\times 2$-matrix-valued differential
expression of order $n+1$ such that $[P_\s ,L]$ is an
operator of
multiplication. Writing
$$P_\s =\sum_{\ell=0}^{n+1} C_{n+1-\ell}(x) L^\ell, \lb{P}$$
and utilizing that the commutator $[C_j,L]$ may be written as
$D_j(x)L+E_j(x),$ the condition that $[P_\s ,L]$ is an
operator of
multiplication yields $D_{j+1}(x)=-E_j(x)$ for $j=0,...,n,$
$D_0=0,$ and
$[P_\s ,L]=E_{n+1}(x).$ Note that $C_j(x)$ can be expressed as
$$C_j(x)=k_j(x)I+v_j(x)J+W_j(x), \quad j=0,\dots,n+1,$$
where the $k_j(x)$ and $v_j(x)$ are scalar-valued and the
matrices
$W_j(x)$ have vanishing diagonal elements. Then ($'=d/dx$)
$$D_j=2W_j \text{ and }
 E_j=-W_j Q-QW_j+v_j'I-k_j'J+2v_j JQ-JW_j'.$$
Since $W_j Q+QW_j$
is a multiple of the identity matrix $I$, $D_{j+1}+E_j=0$
shows that
$k_j'=0$
for all $j=0,\dots,n+1,$ and that the following recursion
relation
holds,
\begin{align} \lb{rr}
 W_0&=0,  \\
 v_j'I&=W_j Q+QW_j, \quad
 W_{j+1}=\frac12 J(W_j'-2v_j Q), \quad j=0,\dots,n+1.  \no
\end{align}
One concludes
$$[P_\s ,L]=2v_{n+1}JQ-JW_{n+1}'.$$
Next, let
\begin{align*}
 K_\s (z)&=\sum_{j=0}^{n+1} k_{n+1-j} z^j, \\
 V_\s (z,x)&=\sum_{j=0}^{n+1}v_{n+1-j}(x) z^j, \\
 \calW_\s (z,x)&=\sum_{j=1}^{n+1} W_{n+1-j}(x) z^j,
\end{align*}
so that the recursion relation \eqref{rr} becomes
\begin{align*}
 V'_\s (z,x)I&=\calW_\s (z,x)Q(x)+Q(x)\calW_\s (z,x),\\
 \calW'_\s (z,x)&=2V_\s (z,x)Q(x)-2zJ\calW_\s (z,x)+J[P_\s ,L].
\end{align*}
Hence,
\begin{equation} \label{11021}
[P_\s ,L]=2z\calW_\s (z,x)+2V_\s (z,x)JQ(x)-J\calW'_\s (z,x).
\end{equation}
If $P_\s $ is of order $n+1$ we define the $n^{\rm th}$
order AKNS
equations by
$${\rm AKNS}_n(Q)=Q_t-[P_\s ,L]=0.$$
The first few of these equations are
\begin{align*}
 Q_t &= -v_0 Q'+2c_1 JQ, \no\\
 Q_t &= -\frac{v_0}2 J(Q''-2Q^3)-c_1Q'+2c_2 JQ, \\
 Q_t &= \frac{v_0}4 (Q'''-6Q^2Q')-c_1J(Q''-2Q^3)-c_2Q'
+2c_3 JQ, \no\\
 & \text{etc.}, \no
\end{align*}
where $v_0$ and $c_1,c_2,...$ are arbitrary integration
constants. Upon
rescaling the $t$ variable one may choose $v_0=1.$ In terms
of the AKNS
pair $(p,q)$ the homogeneous versions (that is, $c_1=c_2=
\dots =0$) of
these equations read
\begin{align*}
\begin{pmatrix} p_t\\q_t \end{pmatrix}
&=-v_0\begin{pmatrix} p_x\\q_x\end{pmatrix}, \notag \\
\begin{pmatrix} p_t\\q_t \end{pmatrix}
&=\frac{iv_0}2 \begin{pmatrix} p_{xx}-2p^2q \\
-q_{xx}+2 pq^2\end{pmatrix}, \notag \\
\begin{pmatrix} p_t\\q_t \end{pmatrix}
&=\frac{iv_0}4 \begin{pmatrix} p_{xxx}-6pqp_x \\
-q_{xxx}+6 pq q_x\end{pmatrix}, \notag \\
& \text{etc.} \no
\end{align*}

We also mention an interesting scale invariance of the
AKNS equations.
Suppose $Q$ satisfies one of the AKNS equations, that is,
${\rm
AKNS}_n(Q)=0.$ Suppose $a\neq0$ and
$$A=\begin{pmatrix}a&0\\0&a^{-1}\end{pmatrix}.$$
Then $\breve Q=AQA$ also satisfies
${\rm AKNS}_n(\breve Q)=0.$ We omit
the straightforward proof which can be found, for
instance, in
\cite{GR98}.

In the particular case of the nonlinear Schr\"{o}dinger
(NS) hierarchy,
where $p(x,t)=\pm \overline{q(x,t)}$ the matrix $A$ is
unimodular, that
is, $|a|=1.$

Note that the KdV hierarchy as well as the modified
Korteweg-de Vries
(mKdV) hierarchy are contained in the AKNS hierarchy. In
fact, setting
all integration constants $c_{2\ell+1}$ equal to zero,
the $n^{\rm th}$
KdV equation is obtained from the $(2n)^{\rm th}$ AKNS
system
by the
constraint $p(x,t) = 1,$ while the $n^{\rm th}$ mKdV
equation is
obtained from the $(2n)^{\rm th}$ AKNS system by the
constraint
$p(x,t)=\pm q(x,t).$

Just as in the KdV case one makes the following observation:
Suppose a
polynomial $V_\s $ whose coefficients are scalar functions
and a
polynomial $\calW_\s $ whose coefficients are $2\times2$
matrix-valued
functions with zero diagonal to be given. Furthermore,
assume
that
$V'_\s (z,x)I =\calW_\s (z,x)Q(x)+Q(x)\calW_\s (z,x)$
and that
$2z\calW_\s (z,x) +2V_\s (z,x)JQ(x)-J\calW'_\s (z,x)$ is
independent of
$z.$ Then the coefficients of $V_\s (z,x)$ and
$\calW_\s (z,x)$
define a
differential expression $P_\s $ which satisfies
\eqref{11021}.

Next, suppose that $[P_\s ,L]=0$ and, without loss of
generality,
$v_0\neq0.$ Then,
\begin{align} \label{17021}
 (P_\s -K_\s (L))^2&=(JV_\s (L,x)+\calW_\s (L,x))^2 \\
 &=\calW_\s (L,x)^2-V_\s (L,x)^2 I \no \\
 &=\sum_{m=0}^{2n+2}a_m(x)L^m, \no
\end{align}
where
$$a_m(x)I=\sum_{\ell+k=m}(W_{n+1-\ell}(x)W_{n+1-k}(x)-
 v_{n+1-\ell}(x) v_{n+1-k}(x)I)$$
is a multiple of the identity matrix. Moreover,
differentiating $(JV_\s
(z,x)+\calW_\s (z,x))^2$ with respect to $x$ yields
$$-2V_\s (z,x)V'_\s (z,x)I+\calW_\s (z,x)\calW'_\s (z,x)
 +\calW'_\s (z,x)\calW_\s (z,x)=0,$$
using
\begin{align*}
V'_\s (z,x)I&=\calW_\s (z,x)Q(x)+Q(x)\calW_\s (z,x),\\
\calW'_\s (z,x)&=2V_\s (z,x)Q-2zJ\calW_\s (z,x),
\text{ and}\\
0&=J\calW_\s (z,x) +\calW_\s (z,x)J.
\end{align*}
Hence the coefficients $a_m(x)$ in \eqref{17021} may be
interpreted as
constant scalars. Since $a_{2n+2}=-v_0^2$ one infers
$(P_\s -K_\s (L))^2
+\R(L)=0,$ where $\R$ is a polynomial of degree $2n+2$ with
complex
coefficients. Hence, if $[P_\s ,L]=0$, then the pair
$(P_\s ,L)$ is
associated with a hyperelliptic curve of (arithmetic)
genus $n$.

Next assume that $\F :\bbC^2\to\bbC_\infty$ is a polynomial
of degree
$n$ in its first variable with scalar meromorphic
coefficients. Denote
the leading coefficient by $-iq(x)$ and let $p(x)$ be another
nonzero
meromorphic function. Defining
$$V_\s (z,x)=\frac{-1}{2q(x)}(\F '(z,x)+2iz \F (z,x))$$
and
$$\calW_\s (z,x)=\frac{i}{q(x)} \begin{pmatrix} 0& q(x)
\F (z,x)\\
 V_\s '(z,x)+p(x)\F (z,x)&0\end{pmatrix},$$
this implies $V_\s 'I=\calW_\s  Q+Q\calW_\s $ with $Q$
given as in
\eqref{18021}. Moreover, $V_\s ^2I-\calW_\s ^2=\R(z,x)I,$
where the
scalar $\R(z,x)$ is given by
\begin{align} \label{18022}
\R(z,x)=&\frac{1}{4q(x)^2}\bigl(\F '(z,x)^2-
2\F (z,x)\F ''(z,x)
+4(p(x)q(x)-z^2) \F (z,x)^2\bigr) \\ &+\frac{q'(x)}
{4q(x)^3}\bigl(2\F
(z,x)\F '(z,x)+4iz\F (z,x)^2\bigr).
\no
\end{align}
If this is constant then differentiation with respect to
$x$ yields
\begin{align*}
\calW_\s (z,x)(\calW'_\s (z,x)&-2V_\s (z,x)Q(x))\\
&+(\calW'_\s (z,x)-2V_\s (z,x)Q(x))\calW_\s (z,x)=0.
\end{align*}
Since $\calW'_\s (z,x)-2V_\s (z,x)Q(x)$ has zero diagonal
elements, this
may be considered, for each fixed $x,$ a linear homogeneous
equation for
the off-diagonal elements of $\calW'_\s (z,x)-
2V_\s (z,x)Q(x).$
This
equation has a one-dimensional space of solutions, in fact,
$\calW'_\s
(z,x)-2V_\s (z,x)Q(x)=r(x)J\calW_\s (z,x).$ Comparing the
leading
coefficients yields $r(x)=-2z$ and hence we have shown that
$\calW'_\s
(z,x)=
-2zJ\calW_\s (z,x)+2V_\s (z,x)Q(x).$ Summarizing, the
following theorem holds.

\begin{theorem} \label{agaknscon}
If $\F :\bbC^2\to\bbC_\infty$ is a polynomial of degree $n$
in the first
variable and meromorphic in the second, and if the expression
$\R(z,x)$
in \eqref{18022} is independent of $x,$ then there exists a
$2\times 2$
matrix-valued differential expression $P_\s $ of order $n+1$
with
leading coefficient $J^{n+2}$ such that $[P_\s ,L]=0.$
\end{theorem}

We are now ready to define the term algebro-geometric in the
context of the AKNS hierarchy.

\begin{definition} \label{agakns}
Suppose $Q$ is a $2\times 2$ matrix-valued meromorphic
function and
let $L$ be the differential expression $L=Jd/dx+Q.$ Then
$Q$ is called an {\it algebro-geometric AKNS potential} (or
simply
{\it algebro-geometric}) if $Q$ is a solution of some
equation of the stationary AKNS hierarchy.
\end{definition}

Equivalently, $Q$ is algebro-geometric if any one of the
following two
conditions is satisfied.
\begin{enumerate}
 \item [(1)] There exists a differential expression $P_\s $
of order
$n+1$ with leading coefficient $J^{n+2}$ such that
$[P_\s ,L]=0.$
 \item [(2)] There exists a function $\F:\bbC^2\to\bbC_\infty$
which is a
polynomial in the first variable, meromorphic in the second,
such that
the expression $\R(z,x)$ in \eqref{18022} does not depend
on $x.$
\end{enumerate}

\subsection{Periodic AKNS Potentials.} \label{paknsp}

Throughout this section we will assume that
$p,q\in C^2(\bbR)$ are
complex-valued periodic functions of period $\Omega>0.$
Matrix-valued solutions $Y$ of
the initial value problem $(LY)(x)=zY(x)$ and
$Y(x_0)=Y_0$ are
denoted by
$\Phi(z,\cdot,x_0,Y_0),$ which is also the unique solution
of the
integral equation
\begin{equation} \label{3.4}
Y(x)=\exp(z(x-x_0)J)Y_0+\int_{x_0}^x \e^{z(x-x')J}
JQ(x') Y(x') dx'.
\end{equation}

In contrast to the Sturm-Liouville case, the Volterra
integral equation
\eqref{3.4} is not suitable to determine the asymptotic
behavior of
solutions as $z$ tends to infinity. To circumvent this
difficulty one
can follow Marchenko's approach in \cite{Ma86},
Section~1.4, to
obtain the asymptotic expansion
\begin{align}
\Phi(z,x_0 + \Omega,x_0,I)=&
\begin{pmatrix} \e^{-iz\Omega}&0\\0&\e^{iz\Omega}
\end{pmatrix}
+\frac{1}{2iz}\begin{pmatrix}\beta\e^{-iz\Omega}&2q(x_0)
\sin(z\Omega) \notag\\
2p(x_0)\sin(z\Omega) &-\beta\e^{iz\Omega}\end{pmatrix}\\
&+O(\e^{|\Im(z)| \Omega} z^{-2}), \label{17025}
\end{align}
where
$$\beta=\int_{x_0}^{x_0+\Omega} p(t)q(t) dt,$$
provided $p, q \in C^2(\bbR).$ From this result we infer
that the entries of $\Phi(\cdot,x_0+\Omega,x_0,I),$ which
are entire
functions, have order of growth equal to one whenever
$q(x_0)$
and $p(x_0)$ are nonzero.

As in the scalar case we introduce ${\calT}_\Omega(z),$ the
restriction
of the translation operator $Y\mapsto Y(\cdot+\Omega)$
to the
two-dimensional space of solutions ${\calY}(z)$ of
$(LY)(x)=zY(x).$ The
Floquet multipliers are then determined as solutions of
$$\rho^2 - \rho\tr ({\calT}_\Omega(z)) + 1=0.$$
They are degenerate if and only if $\rho(z)^2=1.$ The
points $z$ where
this happens are the \hbox{(anti-)}periodic eigenvalues
which we denote
by $E_j,$ $j\in\bbZ.$ Their algebraic multiplicities are
given by
$\ord_{E_j}(\calT_\Omega^2-4)$ and are denoted by
$p(E_j).$ The
asymptotic
behavior
of the \hbox{(anti-)}periodic eigenvalues may be
determined from
\eqref{17025}. It is given by
$$E_{2j}, E_{2j-1}=\frac{j\pi}{\Omega}+O(|j|^{-1}),$$
if the eigenvalues are labeled in accordance with their
algebraic
multiplicity.

The conditional stability set $\mathcal S(L)$ of $L$ is again
$$\mathcal S(L)=\{z\in\bbC\,|-2 \leq
\tr ({\calT}_\Omega(z))\leq 2\}.$$
The spectrum of the maximal operator in $L^2(\bbR)^2$
associated with
$L$ coincides with the conditional stability set
$\mathcal S(L)$ of $L.$
One can also prove that the conditional stability set
$\mathcal S(L)$
consists of a countable number of regular analytic arcs,
the spectral
bands. At most two spectral bands extend to infinity and
at most
finitely many spectral bands are closed arcs. The point
$z$ is a band
edge, that is, an endpoint of a spectral band, if and
only if $(\tr
(\calT_\Omega(z)))^2-4$ has a zero of odd order. For
additional
results on
nonself-adjoint periodic Dirac-type operators, see Tkachenko
\cite{Tk94a}, \cite{Tk97}, \cite{Tk98}.

The boundary value problem defined by the differential
expression
$L$ and the
requirement that the first component of a solution
vanishes at
$x_0$ and
$x_0+\Omega$  will (somewhat artificially but in close
analogy to the
KdV case) be called the Dirichlet problem
for the interval $[x_0,x_0+\Omega].$ The Dirichlet eigenvalues
and their
algebraic multiplicities are given as the zeros and their
multiplicities
of the function
$$g(z,x_0)=(1,0)\Phi(z,x_0+\Omega,x_0,I)(0,1)^t,$$
that is, the entry in the upper right corner of
$\Phi(z,x_0+\Omega,x_0,I).$
The algebraic multiplictiy of $z$ as a Dirichlet eigenvalue
$\mu(x)$ is
denoted by $d(z,x).$ The quantities
$$d_i(z)=\min\{d(z,x)\,|\, x\in\bbR\} \text{ and }
 d_m(z,x)=d(z,x)-d_i(z)$$
will be called the immovable part and the movable part of
the algebraic
multiplicity $d(z,x),$ respectively. The sum $\sum_{z\in\bbC}
d_m(z,x),$
which is independent of $x,$ is called the number of movable
Dirichlet
eigenvalues. Asymptotically, the Dirichlet eigenvalues are
distributed
according to
$$\mu_j(x_0)=\frac{j\pi}{\Omega}+O(|j|^{-1}),
\quad j\in\bbZ.$$

If $q(x)\neq0,$ the function $g(\cdot,x)$ is an entire
function with
order of growth equal to one. Hadamard's factorization
theorem then
implies $g(z,x)=F(z,x) D(z),$ where $F$ comprises the
factors depending
on $x,$ while $D$ contains the factors independent of $x.$
Of course,
$\ord_z(F(\cdot,x))=d_m(z,x)$ and $\ord_z(D)=d_i(z).$

We now turn to the $x$-dependence of the function $g.$ The
following is
the analog of equation \eqref{09121} and is obtained by a
straightforward computation. The function $F(z,\cdot)$
satisfies the
differential equation
\begin{align} \label{3.30}
 &q(x)(F'(z,x)^2-2F(z,x)F''(z,x)+4(p(x)q(x)-z^2)F(z,x)^2) \\
 &+q'(x)(2F(z,x) F'(z,x)+4iz F(z,x)^2)
 =q(x)^3((\tr ({\calT}_\Omega(z)))^2-4)/D(z)^2. \no
\end{align}

Just as in the KdV case we get the following theorem.
\begin{theorem}
There exists an entire function $\ul{R}$ such that $(\tr
({\calT}_\Omega(z)))^2-4=\ul{R}(z)\linebreak[0]D(z)^2.$ In
particular,
$p(z)-2d_i(z)\geq0$ for every $z\in\bbC.$
\end{theorem}

We define the Floquet deficiency of $Q$ (or $L$) as $\df(Q)=
\deg(\ul{R})
\in\bbN_0\cup\{\infty\}.$ Equation \eqref{3.30},
Theorem~\ref{agaknscon}, and the asymptotic distribution of
\hbox{(anti-)}periodic and Dirichlet eigenvalues allow one to
prove the
following central theorem.

\begin{theorem} \label{aknsper}
The following statements are equivalent:
\begin{itemize}
 \item [(1)] $\df(Q)=2n+2.$
 \item [(2)] There are $n$ movable Dirichlet eigenvalues.
 \item [(3)] There exists a $2\times2$ matrix-valued
differential
expression $P_\s $ of order $n+1$ and leading coefficient
$J^{n+2}$
which commutes with $L$. The number $n+1$ is the smallest
integer with
that property. The differential expression $P_\s $ satisfies
$P^2_\s=R_{2n+2}(L) =\prod_{z\in\bbC} (L-z)^{p(z)-2d_i(z)}$
and hence
$Q$ is algebro-geometric.
\end{itemize}
\end{theorem}
We remark that in (3) $R_{2n+2}$ is a constant multiple of
$\ul{R}$.

\subsection{Necessary Conditions for an AKNS Potential to be
 Algebro-Geome\-tric.} \lb{naknspag}

In this section we will prove the following result.
\begin{theorem} \label{t6.3}
Suppose $Q$ is an algebro-geometric AKNS potential. Then
$LY=JY'+QY=zy$
has a meromorphic fundamental system of solutions with
respect to $x$
for all values of the spectral parameter $z\in\bbC.$
\end{theorem}

A proof of this theorem along the lines of the proof of
Theorem~\ref{mt1} fails since one can only show that the
exponents of a
singularity are half integers rather than integers.
However, such
reasoning can be used to prove the following result.

\begin{theorem} \label{t05051}
Suppose $Q$ is a meromorphic potential coefficient of $L.$
Then the
equation $(LY)(x)=zY(x)$ has a fundamental system of solutions
meromorphic with respect to $x$ for all values of the spectral
parameter
$z\in\bbC$ whenever this holds for infinitely many values
of $z.$
\end{theorem}

The rest of this section is devoted to a proof of
Theorem \ref{t6.3}.
Suppose $Q$ is an algebro-geometric AKNS potential. According
to the
characterizations following Definition \ref{agakns} there is
a function
$\F:\bbC^2\to\bbC_\infty$ which is a polynomial of degree
$n$ in the
first variable, meromorphic in the second, such that the
expression
$\R(z,x)$ in \eqref{18022} does not depend on $x$. With the
aid of $\F$
we define (as above)
\begin{align*}
 V_\s (z,x)&=\frac{-1}{2q(x)}(\F '(z,x)+2iz \F (z,x)),\\
 \calW_\s (z,x)&=\frac{i}{q(x)} \begin{pmatrix} 0& q(x)
\F (z,x)\\
 V_\s '(z,x)+p(x)\F (z,x)&0\end{pmatrix},
\end{align*}
and
$$P_\s =JV_\s (L,x)+\calW_\s (L,x).$$
The pair $(P_\s ,L)$ is associated with the hyperelliptic curve
$$\calK_n=\{(z,w):w^2+\R(z)=0\}$$
of arithmetic genus $n$, where
$$\R(z)I=V_\s (z,x)^2I-\calW_\s (z,x)^2=
\prod_{m=1}^{2n+2}(z-E_m)I.$$
We emphasize that the points $E_m$ are not necessarily distinct.

If $\Psi=(\psi_1,\psi_2)^t$ is a common solution of
$L\Psi=z\Psi$ and
$P_\s \Psi=w\Psi$, then, considering the first component of
$P_\s \Psi$,
we find
\begin{align*}
w\psi_1&=(P_\s \Psi)_1=[(JV_\s (z,x)+\calW_\s (z,x))\Psi]_1
 =iV_\s \psi_1+\calW_{\s,1,2}\psi_2\\
 &=(iV_\s +\calW_{\s,1,2}\phi)\psi_1,
\end{align*}
defining $\phi=\psi_2/\psi_1$. We now revert this process and
define the
meromorphic function $\phi$ on $\calK_n\times\bbC$ by
$$\phi((z,w),x)=\frac{w-iV_\s (z,x)}{\calW_{\s,1,2}(z,x)}
 =\frac{\calW_{\s,2,1}}{w+iV_\s (z,x)}.$$
We remark that $\phi$ can be extended to a meromorphic
function on the
compactification (projective closure) of the affine curve
$\calK_n.$
This compactification is obtained by joining two points
(the points at
infinity) to $\calK_n.$

Next we define
\begin{align}
\psi_1((z,w),x,x_0) &= \exp \bigg(\int_{x_0}^x
(-iz+q(x')\phi((z,w),x')) dx'\bigg), \label{6.3} \\
\psi_2((z,w),x,x_0) &= \phi((z,w),x)\psi_1((z,w),x,x_0)
\end{align}
where the simple Jordan arc from $x_0$ to $x$ in
\eqref{6.3} avoids
poles of $q$ and $\phi.$ One verifies with the help
of $V_\s 'I
=\calW_\s
Q+Q\calW_\s $ and $\calW_\s '=2V_\s Q-2zJ\calW_\s $ that
$$\phi'((z,w),x)=p(x)-q(x)\phi((z,w),x)^2+2iz\phi((z,w),x).$$
>From this one obtains that
$$\Psi((z,w),x,x_0)=\begin{pmatrix}
\psi_1((z,w),x,x_0) \\ \psi_2((z,w),x,x_0)\end{pmatrix}$$
is a common solution of $L\Psi=z\Psi$ and $P_\s \Psi=w\Psi$.

Let $\mu_1(x_0),...,\mu_n(x_0)$ denote the zeros of
$\F(\cdot,x_0)$. One
then observes that the two branches $\Psi_{\pm}(z,\cdot,x_0)
=\Psi((z,\pm
w),\cdot,x_0)$ of the function $\Psi((z,w),\cdot,x_0)$
represent a
fundamental system of solutions of $Ly=zy$ for all
complex numbers $z$
different from $E_1,...,E_{2n+2},\mu_1(x_0),...,\mu_n(x_0)$,
since
\begin{equation} \label{6.8}
W(\Psi_-(z,x,x_0), \Psi_+(z,x,x_0))=\frac{2w}
{\calW_{\s,1,2}(z,x_0)}
=\frac{-2iw}{\F (z,x_0)},
\end{equation}
where $W(f,g)$ denotes the determinant of the two columns
$f$ and $g$.

In the special case where $\calK_n$ is nonsingular, that is,
when the
points $E_m$ are pairwise distinct, the explicit
representation of
$\Psi((z,w),x,x_0)$ in terms of the Riemann theta function
associated
with $\calK_n$ immediately proves that $\Psi_{\pm}(z,x,x_0)$
are
meromorphic with respect to $x\in\bbC$ for all $z\in
\bbC\backslash\{E_1,...,E_{2n+2},\mu_1(x_0),...,\mu_n(x_0)\}.$
A detailed account of this theta function representation
can be found,
for instance, in Theorem~3.5 of \cite{GR98}. In the
following we
demonstrate how to use gauge transformations to reduce the
case of
singular curves $\calK_n$ to nonsingular ones.

Let $Q$ be meromorphic on $\bbC$ and introduce
$A(z,x)=(Q(x)+zI)J$ which
turns $LY=zY$ into $Y'+AY=0.$ Then consider the gauge
transformation
$$\tilde \Psi(z,x)=\Gamma(z,x)\Psi(z,x).$$
If
$$\tilde A(z,x)=\Gamma(z,x)A(z,x)\Gamma(z,x)^{-1}
-\Gamma'(z,x)\Gamma(z,x)^{-1},$$
then $\tilde A(z,x)=(\tilde Q(x)+zI)J,$ where $\tilde Q$
has zero
diagonal elements. Moreover, $\tilde \Psi'
+\tilde A\tilde\Psi=0,$
that is,
$(Jd/dx+\tilde Q)\tilde\Psi=z\tilde\Psi.$ Next we make the
following
explicit choice suggested by Konopelchenko \cite{Ko82}. Let
$\tilde z\in\bbC$
be fixed, $\Psi^{(0)}(\tilde z,\cdot)=(\psi^{(0)}_1(\tilde
z,\cdot),\psi^{(0)}_2(\tilde z,\cdot))^t$ be any solution of
$Ly=\tilde
zy,$ and introduce
$$\phi^{(0)}(\tilde z,x)
 =\psi^{(0)}_2(\tilde z,x)/\psi^{(0)}_1(\tilde z,x).$$
Then one defines
$$\Gamma(z,x)=\frac12 \begin{pmatrix}
 2(z-\tilde z)-iq(x)\phi^{(0)}(\tilde z,x)& iq(x)\\
 i\phi^{(0)}(\tilde z,x)&-i\end{pmatrix}$$
for $z\in \bbC\backslash\{\tilde z\}.$

The upper right entry $G_{1,2}(z,x,x')$ of the Green's matrix
of $L$ is
given by
$$G_{1,2}(z,x,x')=i\frac{\psi_{+,1}(z,x,x_0)
\psi_{-,1}(z,x',x_0)}
 {W(\Psi_-(z,\cdot,x_0),\Psi_+(z,\cdot,x_0))},
\quad x \geq x'.$$
We want to evaluate $G_{1,2}(z,x,x')$ on its diagonal
(i.e., where
$x=x'$). Since $\psi_{+,1}(z,x_0,x_0)\psi_{-,1}(z,x_0,x_0)=1$
we obtain
from \eqref{6.3} and \eqref{6.8}
$$G_{1,2}(z,x,x)=i\frac{\calW_{\s,1,2}(z,x)}{2w}=
\frac{-\F (z,x)}{2w}.$$

Next note that
$$W(\tilde\Psi_-(z,\cdot),\tilde\Psi_+(z,\cdot))
 =\det(\Gamma(z,x))W(\Psi_-(z,\cdot),\Psi_+(z,\cdot))$$
and that $\det(\Gamma(z,x))=-i(z-\tilde z)/2$. With the
help of this
fact and some computations one finds that the upper right
entry $\tilde
G_{1,2}(z,x,x)$ of the Green's matrix of $Jd/dx+\tilde Q$ is
$$\tilde G_{1,2}(z,x,x)
=i\frac{\tilde\psi_{+,1}(z,x)\tilde\psi_{-,1}(z,x)} {W(\tilde
\Psi_-(z,\cdot),\tilde \Psi_+(z,\cdot))} =\frac{-\tF(z,x)}
{2(z-\tilde z)w},$$ where $\tF(\cdot,x)$ is a polynomial of
degree $n+1$
with leading coefficient $-i\tilde q(x).$ Moreover, $\tF(z,x)$
satisfies
\begin{align}
&\tilde q(2\tF\tF''-\tF^{\prime2} +4(z^2-\tilde p\tilde q)\tF^2)
-\tilde q'(2\tF\tF'+4iz \tF^2)\notag\\
=&-4 \tilde q^3 (z-\tilde z)^2 \R(z). \label{19021}
\end{align}
Hence $\tilde Q$ is an algebro-geometric AKNS potential.

Now suppose that $(\tilde z,0)$ is a singular point of
$\calK_n,$ that
is, that $\tilde z$ is a zero of $\R$ of order $r\geq2.$
Choose
$$\phi^{(0)}(\tilde z,x)
 =\frac{-iV_\s (\tilde z,x)}{\calW_{\s,1,2}(\tilde z,x)}.$$
Then one may show that $\tF(\cdot,x)$ has a zero of order at
least $2$
at $\tilde z,$ that is,
$$\tF(z,x)=(z-\tilde z)^s \hF(z,x)$$
for some $s\geq 2$ and some polynomial $\hF(\cdot,x)$ of degree
$\hat{n}=n+1-s$. From \eqref{19021} one obtains
$$\tilde q(2\hF\hF''-\hF^{\prime2}+
4(z^2-\tilde p\tilde q)\hF^2)
 -\tilde q'(2\hF\hF' +4iz \hF^2)=-4 \tilde q^3 \hR(z),$$
where
$$\hR(z)=(z-\tilde z)^{2-2s} \R(z)$$
is a polynomial in $z$ of degree $2n-2s+4\in(0,2n+2).$ This
proves that
$\tilde Q$ is associated with the curve
$$\calK_{\hat n}=\{(z,w): w^2+(z-\tilde z)^{2-2s}\R(z)=0\}.$$

Our choice of $\phi^{(0)}(\tilde z,x)$ led to a curve
$\calK_{\hat n}$
which is less singular at $(\tilde z,0)$ than $\calK_n,$
without
changing the local structure of the curve elsewhere. By
iterating this
procedure one ends up with a curve which is nonsingular at
$(\tilde
z,0).$ Repeating the procedure for each singular point of
$\calK_n$ then
results in a nonsingular curve and a corresponding
Baker-Akhiezer
function $\hat\Psi((z,w),x,x_0)$ meromorphic with respect to
$x\in\bbC$
using its standard theta function representation (see, e.g.,
\cite{GR98}). Since $\phi^{(0)} =-iV_\s /\calW_{\s,1,2}$ is
meromorphic,
the gauge transformations and their inverses map meromorphic
functions
to meromorphic functions. Combining these results proves
Theorem~\ref{t6.3}.

A systematic account of the effect of Darboux-type
transformations on
hyperelliptic curves in connection with the KdV, AKNS, and Toda
hierarchies will be presented in \cite{GH98}.

\subsection{Sufficient Conditions for an AKNS Potential
to be
 Algebro-Geome\-tric -- A Characterization of Elliptic
Stationary
AKNS Solutions.} \lb{saknspag}

\begin{definition} \label{defpicaknspot}
Let $p,q$ be elliptic functions with a common period lattice.
Then
$Q=\left(\begin{smallmatrix} 0& -iq \\ ip & 0
\end{smallmatrix}\right)$ is called a {\it Picard-AKNS
potential} if
the equation $J\psi'(x)+Q(x)\psi(x)=z\psi(x)$ has a
meromorphic
fundamental system of solutions with respect to $x$ for all
values
of the spectral parameter $z\in\bbC.$
\end{definition}

We note
that according to Theorem~\ref{t05051} it is sufficient to
show the
existence of a meromorphic fundamental system for infinitely
many values
of $z$ in order to prove that $Q$ is a Picard-AKNS potential.

Just as in the KdV case the following theorem characterizes
all elliptic
algebro-geometric AKNS potentials.
\begin{theorem}\label{t5.4}
Let $Q=\left(\begin{smallmatrix} 0& -iq \\ ip & 0
\end{smallmatrix}\right)$ with $p,q$ elliptic functions with a
common
period lattice. Then $Q$ is an elliptic algebro-geometric AKNS
potential if and only if it is a Picard-AKNS potential.
\end{theorem}
\begin{proof}[Sketch of proof]
The necessity of the criterion is the content of
Theorem~\ref{t6.3}. The
sufficiency follows from Picard's Theorem~\ref{picard} and
Theorem~\ref{aknsper} in the same way as in the KdV case.
\end{proof}

The transformation
$$\tilde\Psi(x)
 =\begin{pmatrix} \e^{ax+b}&0\\0&\e^{-(ax+b)}
\end{pmatrix}\Psi(x)$$
allows one to prove the following corollary, which slightly
extends
the class of algebro-geometric AKNS potentials $Q$ considered
thus far.
Such cases have recently
been considered by Smirnov \cite{Sm95}.

\begin{corollary}\label{c5.5}
Suppose
\begin{equation}
Q(x)=\begin{pmatrix} 0&-i q(x) \e^{-2(ax+b)}
\\ip(x)\e^{2(ax+b)}&0
\end{pmatrix},
\end{equation}
where $a,b\in\bbC$ and $p,q$ are elliptic functions with
a common period
lattice. Then $Q$ is an algebro-geometric AKNS potential if
$J\Psi'(x)+Q(x)\Psi(x)=z\Psi(x)$ has a meromorphic fundamental
system
of solutions with respect to $x$ for all values of the
spectral parameter $z\in\bbC.$
\end{corollary}

\subsection{Examples.} \label{ex}
\setcounter{equation}{0}

With the exception of the studies by Christiansen, Eilbeck,
Enol's\-kii,
and Kostov \cite{CEEK95} and Smirnov \cite{Sm95}, \cite{Sm96a},
\cite{Sm97}, not too many examples of elliptic solutions
$(p,q)$ of the
AKNS hierarchy associated with higher (arithmetic) genus
curves of the
type $w^2+\R(z)=0$ have been worked out in detail. The genus
$n=1$ case
has been considered, for example, by Its \cite{It81} and
Pavlov
\cite{Pa87}. Moreover, examples for low genus $n$ for special
cases such
as the nonlinear Schr\"odinger and mKdV equation are
considered, for
instance, in \cite{AIK90}, \cite{BBM86a}, \cite{LT90},
\cite{Me87},
\cite{MS93}, \cite{OB90}, \cite{Sm95a}. Examples related
to equations of
the sine-Gordon-type are discussed in \cite{Sm91},
\cite{Sm97a},
\cite{Sm97b}. The following examples in
\eqref{e7.1} -- \eqref{e7.4} are
algebro-geometric AKNS potentials as can be proved using
the Frobenius
method. For details on this procedure see \cite{GW98}.

\begin{align}
Q(x)&=in(\zeta(x)-\zeta(x-\omega_2)-\zeta(\omega_2))
\begin{pmatrix}0&-1\\1&0\end{pmatrix},
\quad n\in\bbN,\label{e7.1}\\
Q(x)&=\begin{pmatrix}0&-in(n+1)\wp(x)\\i&0\end{pmatrix},
\quad n\in\bbN, \label{e7.2}\\
Q(x)&=\begin{pmatrix}0&i\wp'(x-\omega_2)/(2e_1)\\
 -3i\wp'(x)/(2e_1)&0\end{pmatrix},\quad e_2=0, \label{e7.3}\\
Q(x)&=\begin{pmatrix}0&i\wp(x-\omega_2)/e_1^2\\
2i(\wp''(x)-e_1^2)/3&0\end{pmatrix},\quad e_2=0.\label{e7.4}
\end{align}

Incidentally, if $p=1,$ as in Example~\eqref{e7.2} then
$J\Psi'+Q\Psi=z\Psi$ is equivalent to the scalar equation
$\psi_2''-q\psi_2=-z^2\psi_2,$ where $\Psi=(\psi_1,\psi_2)^t$
and
$\psi_1=\psi_2'-iz \psi_2.$ Therefore, if $-q$ is an elliptic
algebro-geometric potential of the KdV hierarchy then, by
Theorem
\ref{t3.12}, $\psi_2$ is meromorphic for all values of $z.$
Hence $\Psi$
is meromorphic for all values of $z$ and therefore $Q$ is a
Picard-AKNS
and hence an algebro-geometric AKNS potential. Conversely, if
$Q$ is an
algebro-geometric AKNS potential with $p=1$ then $-q$ is an
algebro-geometric potential of the KdV hierarchy. In particular,
$q(x)=n(n+1)\wp(x)$ is again the class of Lam\'e potentials
associated
with the KdV hierarchy and hence a special case of the material
discussed in Section~\ref{pickdvp}.

\bibliographystyle{amsplain}

\end{document}